\documentclass[12pt]{iopart}
\usepackage[T1]{fontenc}
\usepackage{graphicx}
\expandafter\let\csname equation*\endcsname\relax
\expandafter\let\csname endequation*\endcsname\relax
\usepackage{amsmath}
\usepackage{amssymb}
\usepackage{epsfig}
\usepackage{longtable}
\usepackage{subfigure}
\usepackage{multirow}
\usepackage{color}
\usepackage{epstopdf}  
\usepackage{ifpdf}
\ifpdf
	\usepackage{epstopdf}
\fi

\newcommand{\IYZ}[1]{\textcolor{green}{{\bf IYZ: #1 }}}

\begin{document}

\bibliographystyle{iopart-num}

\title{{\rm\small\hfill }\\
   Introduction to the Fifth-rung Density Functional Approximations: Concept, Formulation, and Applications}

\author{Igor Ying Zhang$^1$ \footnote{igor\_zhangying@fudan.edu.cn} 
and Xinguo Ren$^{2,3}$ \footnote{renxg@iphy.ac.cn}}
\address{$^1$ Collaborative Innovation Center of Chemistry for Energy Materials, Shanghai, 
 Key Laboratory of Molecular Catalysis and Innovative Materials,
 MOE Key Laboratory of Computational Physical Sciences, 
 Shanghai Key Laboratory of Bioactive Small Molecules,
 Department of Chemistry, Fudan University, Shanghai 200433, China}
\address{$^2$  Institute of Physics, Chinese Academy of Sciences, Beijing 100190, China}
\address{$^3$  Songshan Lake Materials Laboratory, Dongguan 523808, Guangdong, China}
\pagestyle{plain}
\pagenumbering{arabic}


\newcommand{\kpe}{\mathbf{k}\!\cdot\!\mathbf{p}\,}
\newcommand{\Kpe}{\mathbf{K}\!\cdot\!\mathbf{p}\,}
\newcommand{\bfr}{ {\bf r}} 
\newcommand{\bfrp}{ {\bf r'}} 
\newcommand{\tp}{ {t^\prime}} 
\newcommand{\bfrpp}{ {\bf r^{\prime\prime}}} 
\newcommand{\bfR}{ {\bf R}} 
\newcommand{\bfRp}{ {\bf R'}} 
\newcommand{\bfG}{ {\bf G}} 
\newcommand{\bfGp}{ {\bf G'}} 
\newcommand{\bfzero}{ {\bf 0}} 
\newcommand{\bfq}{ {\bf q}} 
\newcommand{\bfp}{ {\bf p}} 
\newcommand{\bfk}{ {\bf k}} 
\newcommand{\dv}{ {\Delta \hat{v}}} 
\newcommand{\sigmap}{\sigma^\prime} 
\newcommand{\omegap}{\omega^\prime} 
\newcommand{\omegapp}{\omega^{\prime\prime}} 
\newcommand{\bracketm}[1]{\ensuremath{\langle #1   \rangle}}
\newcommand{\bracketw}[2]{\ensuremath{\langle #1 | #2  \rangle}}
\newcommand{\bracket}[3]{\ensuremath{\langle #1 | #2 | #3 \rangle}}
\newcommand{\ket}[1]{\ensuremath{| #1 \rangle}}
\newcommand{\GnWn}{\ensuremath{G^\text{0}W^\text{0}}\,}
\newcommand{\tn}[1]{\textnormal{#1}}
\newcommand{\f}[1]{\footnotemark[#1]}
\newcommand{\mc}[2]{\multicolumn{1}{#1}{#2}}
\newcommand{\mcs}[3]{\multicolumn{#1}{#2}{#3}}
\newcommand{\mcc}[1]{\multicolumn{1}{c}{#1}}
\newcommand{\refeq}[1]{(\ref{#1})} 
\newcommand{\refcite}[1]{Ref.~\cite{#1}} 
\newcommand{\refsec}[1]{Sec.~\ref{#1}} 
\newcommand\opd{d}
\newcommand\im{i}

\begin{abstract}
        The widespread use of (generalized) Kohn-Sham density functional theory (KS-DFT) lies in the fact that hierarchical sets of approximations of
        the exchange-correlation (XC) energy functional can be designed, offering versatile choices to satisfy different levels of accuracy needs. 
        The XC functionals standing on the fifth (top) rung of the Jacob's ladder incorporate the information of unoccupied Kohn-Sham orbitals, and by doing so can describe seamlessly non-local electron correlations that the lower-rung functionals fail to capture. 
        The doubly hybrid approximations (DHAs) and random phase approximation (RPA) based methods are two representative classes
        of fifth-rung functionals that have been under active development over the past two decades. 
        In this review, we recapitulate the basic concepts of DHAs and RPA, derive their underlying theoretical formulation 
     from the perspective of adiabatic-connection fluctuation-dissipation theory, and describe the implementation 
     algorithms based on the resolution-of-identity  technique within an atomic-orbital basis-set framework. Illustrating examples of
     practical applications of DHAs and RPA are presented, highlighting the usefulness of these functionals in resolving challenging problems
     in computational materials science. The most recent advances in the realms of these two types of functionals are briefly 
     discussed.
\end{abstract}

\section{Introduction}
It is now an undisputed fact that the most widely used method for the first-principles electronic-structure calculations
is Kohn-Sham (KS) density functional theory (DFT) \cite{Hohenberg/Kohn:1964,Kohn/Sham:1965}. Although the exact exchange-correlation (XC) energy functional
\index{exchange-correlation energy functional} within the KS-DFT framework is unknown, 
there exist many practical approximations -- density functional approximations (DFAs). 
Due to their excellent trade-off between accuracy and efficiency, these DFAs have been applied with great success 
in chemistry, physics, and materials science. 
However, in order to meet the increasing demand on the simulation accuracy and efficiency, novel functionals with better accuracy at affordable cost are still under active development. 

According to the Jacob's Ladder of DFT proposed by Perdew and Schmidt, the DFAs 
can be categorized into five rungs, depending on the complexity of fundamental ingredients used in the functional
construction \cite{Perdew/Schmidt:2001}. The first rung of Jacob's Ladder is 
the local density approximation (LDA) \cite{Kohn/Sham:1965}, which depends on the electron density only.
On the second rung sit generalized gradient approximations (GGAs),  which require the density and density gradient in its functional formula;
one of the most widely used DFAs in computational materials science is the non-empirical GGA
proposed by Perdew, Burke, and Ernzerhof (PBE) \cite{Perdew/Burke/Ernzerhof:1996}. 
On the third rung of this hierarchy, the so-called meta-GGAs further incorporate a more complex 
ingredient, namely the kinetic energy density, which renders the meta-GGA not an explicit functional of density.
The importance of kinetic energy density for the density functional development was first recognized by Becke in early 1983 \cite{becke_hartreefock_1983}.
The most popular meta-GGAs are M06-L \cite{zhao_new_2006} and SCAN \cite{Sun/Ruzsinszky/Perdew:2015}, primarily used in quantum chemistry 
and computational materials science, respectively.
On the fourth-rung of this hierarchy, the ``exact exchange'' components, or more precisely
the ingredients of occupied KS orbitals,
are further included, resulting in the so-called hybrid (meta)-GGAs.
At present, the most popular DFAs in quantum chemistry are exclusively the fourth-rung functionals,  
including B3LYP \cite{Becke:1993}, M06 \cite{Zhao/Truhlar:2006} and $\omega$B97XD \cite{Chai/Head-Gordon:2008},
to name a few. On the other hand, in computational materials science, the screened hybrid functional 
HSE \cite{Heyd/Scuseria/Ernzerhof:2003} is most widely used.

Nevertheless, no matter how intricate they are, DFAs on the first four rungs described above are constructed with information stemming only from
occupied orbitals. The explicit use of unoccupied orbitals is considered to be the key feature of the DFAs that
stand on the fifth rung -- the highest rung of Jacob's Ladder in DFT.
The fifth-rung DFAs hold the promise to approach the simulation accuracy required in the era of precision chemistry. Doubly hybrid approximation (DHA) and random phase 
approximation (RPA) based functionals are two representative examples of fifth-rung DFAs
that have been under active development over the last two decades. In this review, we will
focus on the basic concept and the current status of these two types of functionals.

\subsection{Doubly Hybrid Approximations}
DHAs are now the leading actor in quantum chemistry. The name of ``doubly hybrid'' was originally used 
by Truhlar and co-workers for their empirical multi-coefficient 
methods of MC3BB and MC3MPW \cite{zhao:2005}, where the DFT total energy is linearly mixed with the total energy of the 
simplest wavefunction-based correlation method, i.e.\ the second-order M\o{}ller-Plesset perturbation theory (MP2).
Later, Grimme introduced another kind of doubly hybrid strategy to connect MP2 and meta-GGAs \cite{Grimme:2006}. 
B2PLYP is the first DHA of this family \cite{Grimme:2006}. Popular B2PLYP-type DHAs include DSD-BLYP \cite{dsd_blyp:2010}, 
$\omega$B97X-2 \cite{chai:2008}, and so forth \cite{goerigk_trip_2019}.

In fact, DHAs can be derived in the generalized KS-DFT framework and viewed as an empirical extension of the G\"orling-Levy
second-order perturbation theory (GL2) \cite{Goerling/Levy:1993} along the adiabatic-connection (AC) approach \cite{Gunnarsson/Lundqvist:1976}.
It was first proposed by Zhang and co-workers in 2009 \cite{zhang:2009A}, which emphasizes on the use of accurate density and orbitals of
KS non-interacting systems for evaluating the total energy of DHAs \cite{zhang:2021A,Zhang/Xu:2014}. XYG3 is the first DHA of this family \cite{zhang:2009A}. 
Popular XYG3-type DHAs includes XYGJ-OS \cite{zhang:2011A}, xDH-PBE0 \cite{zhang:2012A}, $\omega$B97M(2) \cite{head-gordon:2018}, 
xrevDSD-PBEB86-D4 \cite{martin:2019}, and so forth.

The rapid growth of computation capacity has been boosting the development of comprehensive databases for main-group chemistry 
with accurate benchmark reference. Two well-established databases are the MGCDB84 set of the Head-Gordon group with about 5000 
test cases \cite{head-gordon:2018} and the GMTKN55 set of the Grimme group with 1505 test cases \cite{goerigk_look_2017}, respectively.
A serial of recent benchmark works using these databases repeatedly demonstrated that the DHAs present significant and consistent 
improvement over the conventional DFAs on the first four rungs, and are competent for various kinds of chemical interactions of the 
main-group chemistry \cite{martin:2019,zhang:2021A}. 
The XYG3-type DHAs stand out in this round of benchmark according to benchmark results obtained independently by 
different research groups. The top-class performers all belong to the XYG3 family of DHAs, including the XYG7 of the Xu group,
the $\omega$B97M(2) of the Head-Gordon group, and the xrevSDS-PBEB86-D4 of the Martin group. Their performance on main-group chemistry
has been comparable to, or even better than the high-level wavefunction-based methods\cite{zhang:2021A}. 

At present, the mainstream DHAs aforementioned use the second-order perturbation energy (PT2) to capture the electron correlation effect that
involves the unoccupied orbitals. Recently, Santra and co-workers observed further notable improvement 
by introducing the third-order perturbation energy (PT3) in the context of XYG3-type doubly hybrid framework \cite{martin:2021}. 
However, the computational cost increases significantly. Meanwhile, it is well-documented that the perturbation algorithms at
any finite order (PTn) completely fail for describing the strong correlation effect, which is closely related to the so-called static correlation
in quantum chemistry. Therefore, the DHAs including the PT2 and/or PT3 correlation contributions cannot be used to study the bond dissociation
problem and the complex systems involving transition metals. A number of studies have begun to address this limitation of 
current DFAs by introducing some kinds of renormalization algorithms to the standard PT2 with little extra computational costs. 
As will be introduced in the next section, the random-phase approximation (RPA) is another successful fifth-rung DFA in materials science. 
The RPA-type correlation can be viewed as a renormalization of the direct term of PT2. RPA and its variants have also been taken as 
the candidate to replace the PT2 in the DHAs \cite{mezei:2015,grimme/steinmetz:2016,mezei:2017,zhang:2019A,zhang:2021B}.


\subsection{Random Phase Approximation}
RPA \index{Random phase approximation (RPA)} is a fundamental concept that plays a central role in many-body physics. 
The concept was developed by Bohm and Pines \cite{Bohm/Pines:1951,Bohm/Pines:1953} in the early 1950’s in an endeavor to describe the cohesive properties of 
the so-called jellium -- interacting electrons moving in a background of uniform positive charge. Using a Hamiltonian formulation of interacting many-electron
system, Bohm and Pines were able to decouple the collective motion of electrons -- the plasma oscillations -- from their individual motions, a procedure named as RPA. It was soon recognized by Hubbard \cite{Hubbard:1957b} that the original RPA \index{RPA} formulation is equivalent to the infinite summation of ring diagrams \index{ring diagrams} 
from the viewpoint of diagrammatic many-body perturbation theory \cite{Gell-Mann/Brueckner:1957}. Since then, the RPA concept has gone beyond the 
realm of condensed matter physics and significantly influenced all branches of physics.

Although the RPA concept can be applied to any interacting many-particle systems (for its applications in nuclear physics, see e.g., Ref.~\cite{Ring/Schuck:1980}),
the next key step towards applying RPA to real materials was the incorporation of
RPA into the KS-DFT \index{Kohn-Sham density functional theory} framework in the 1970's 
\cite{Langreth/Perdew:1975,Langreth/Perdew:1977,Gunnarsson/Lundqvist:1976}. This formulation turned RPA into a first-principles electronic-structure method, suitable for computing the ground-state energy of real materials. 
Within the KS-DFT framework, RPA can be viewed as a fifth-rung approximation to the exchange-correlation energy functional, according to Jacob's ladder \index{Jacob's ladder} of DFT \cite{Perdew/Schmidt:2001}.
However, the application of RPA to realistic systems was impeded by its high computational cost and the lack of efficient algorithms at the time. The first application of RPA to small molecules only appeared in early 2000's, carried out by Furche \cite{Furche:2001} and by Fuchs and Gonze \cite{Fuchs/Gonze:2002}. Since then, 
RPA has been applied to a variety of systems including
atoms \cite{Jiang/Engel:2007,Hellgren/Barth:2007}, molecules \cite{Furche:2001,Fuchs/Gonze:2002,Toulouse/etal:2009,Zhu/etal:2010,Hesselmann/Goerling:2011b,Ren/etal:2011,Eshuis/Furche:2011}, solids \cite{Harl/Kresse:2008,Harl/Kresse:2009,Harl/Schimka/Kresse:2010,Nguyen/deGironcoli:2009,Lu/Li/Rocca/Galli:2009,Casadei/etal:2012,Casadei/etal:2016,Zhang/Cui/Jiang:2018}, surfaces \cite{Ren/etal:2009,Schimka/etal:2010}, interfaces \cite{Mittendorfer/etal:2011,Olsen/etal:2011},
layered materials \cite{Lebegue/etal:2010}, and defects \cite{Bruneval:2012,Kaltak/Klimes/Kresse:2014}. The consensus arising from these studies 
is that RPA is capable of describing the delicate energy differences in complex chemical environments \cite{Ren/etal:2009,Schimka/etal:2010,Casadei/etal:2012,Zhang/Cui/Jiang:2018,Sengupta/Bates/Ruzsinszky:2018,Cazorla/Gould:2019,Yang/Ren:2022}, the correct asymptotic behavior of van der Waals (vdW) complexes \cite{Dobson/White/Rubio:2006,Ren/etal:2013,Gao/Zhu/Ren:2020} and 
layered materials \cite{Lebegue/etal:2010,Bjoerkman/etal:2012}, and the correct dissociation limit of closed-shell molecules \cite{Furche:2001,Fuchs/Gonze:2002}. 
Evidences show
that RPA can provide unprecedented accuracy compared to lower-rung DFAs at tractable computational cost. As such, RPA is expected to play an increasingly more important role in computational materials science, with the rapid development of more efficient algorithms and the availability of more powerful computing resources. 

In a review paper \cite{Ren/etal:2012b} published in 2012, Ren~\textit{et al.} discussed the history of the RPA concept, 
its formulation as a first-principles method \index{first-principles method}, and its applications in quantum chemistry and computational materials science up to that time. These points were nicely summarized by David Pines in
his recent review paper titled as ``\textit{Emergent behavior in strongly correlated electron systems}" \cite{Pines:2016}. In particular, Pines noted that,
 \begin{quote}
         \textit{
	 Sixty-plus years later, the RPA continues to play a significant role
	 in nuclear physics [66], bosonic field-theory [67], the quarkgluon
	 plasma [68], many-fermion solvable models [69], and
	 especially in computational chemistry and materials science.
	 A recent review by Ren et al [70], to which the interested
	 reader is referred, describes the impact of the RPA in the
	 theoretical chemistry and materials science community, cites
	 some thirty articles that indicate the renewed and widespread
	 interest in the RPA during the period 2001-2011, discusses
	 how it enables one to derive the $1/r^6$ interaction between spatially
	 separated closed shell electron systems, and, shows, in some detail, 
	 how the RPA enables one to go beyond density functional theory in computing
	 ground state energies.}
 \end{quote}
This paragraph highlights the far-reaching impact of RPA \index{RPA} in a variety of fields, in particular in
computational chemistry and materials science. It is worthwhile to mention that several other, 
complementary RPA review papers \cite{Hesselmann/Goerling:2011,Eshuis/Bates/Furche:2012} also appeared around that time, where the theoretical foundation
and applications of RPA are discussed from different perspective. More recent account of RPA of 
review character can be found in Ref.~\cite{Chen/etal:2017}. 
     
In this review, we will first give an account of the theoretical foundation of DHAs and RPA, highlighting their unique role in DFT. 
Computational schemes beyond PT2-based DHAs and RPA will be also briefly discussed in this review.
This is followed by a description of the key algorithm of implementing DHAs and RPA using the resolution-of-identity (RI) technique, 
within an atomic-orbital basis-set framework. 
We then present some prototypical applications illustrating the usefulness of these fifth-rung
functionals particularly in computational materials science. 
Note that, the general performance of DHAs have been extensively discussed for finite molecules, but only a limited number 
of their applications on solids were reported, which are introduced in this review.
Most recent efforts devoted to further developing the theoretical and computational aspects
of DHAs and RPA-based functionals as well as extending their capabilities will be briefly mentioned and commented, before we conclude the review. 

\section{Theoretical foundation of Fifth-rung DFAs}

An interacting $N$-electron system is described by the following Hamiltonian,
   \begin{equation}
       \label{eq:interacting_Hamiltonian}
	   \begin{split}
	   \hat{H}  &= \hat{T} + \hat{V}_\mathrm{ext} + \hat{V}_\mathrm{ee} \\
	       &= -\sum_i^{N} \frac{\nabla^2_i}{2} + \sum_i^{N} v^\mathrm{ext}(\hat{\bfr}_i) + \frac{1}{2} \sum_{i\ne j}^{N} 
	       \frac{1}{|\hat{\bfr}_i - \hat{\bfr}_j|} ,
	   \end{split}
   \end{equation}
where $\hat{T}$, $\hat{V}_\mathrm{ext}$, and $\hat{V}_\mathrm{ee}$ are the Kinetic energy operator, the external potential operator,
and the electron-electron Coulomb interaction operators, respectively. 
The Hartree atomic unit ($\hbar=e=m_e=1$) is used throughout this review. Note that the 
operators $\hat{T}$ and $\hat{V}_\mathrm{ee}$ are universal for any $N$-electron
systems, and a system is completely specified by the external potential,
    \begin{equation}
	    v^\mathrm{ext}(\bfr) = - \sum_{\alpha} \frac{Z_\alpha}{|\bfR_\alpha - \bfr|} \, ,
    \end{equation}
with $Z_\alpha$ and $\bfR_\alpha$ being respectively the nuclear charges and positions of the atoms in the system.
The Hamiltonian in Eq.~\ref{eq:interacting_Hamiltonian} 
cannot be solved, even numerically, for more than a few electrons. 

In KS-DFT\index{KS-DFT} \cite{Hohenberg/Kohn:1964,Kohn/Sham:1965}, the ground-state total energy of the interacting $N$-electron system
is recast to an (implicit) functional of 
the electron density $n(\bfr)$ and is customarily separated into four terms: 
\begin{equation}
 E[n] = T_\text{s}[\psi] + E_\text{ext}[n] + E_\text{H}[n] + E_\text{xc}[n]\, ,
\label{eq:E_KS-DFT}
\end{equation}
where
\begin{equation}
	T_\text{s} =  -\sum_{m,\sigma}^{occ} \left\langle \psi_{m,\sigma} \left| \frac{\nabla^2}{2} \right| \psi_{m,\sigma} \right\rangle 
\end{equation}
	is the kinetic energy of the KS independent-particle system, 
\begin{equation}
	E_\text{ext}[n]=\int d\bfr~ v^\mathrm{ext}(\bfr) n(\bfr) 
\end{equation}
is the external potential energy due to the nuclei, and 
\begin{equation}
	E_\mathrm{H}[n] =  \frac{1}{2} \iint d\bfr d\bfrp~ \frac{n(\bfr)n(\bfrp)}{|\bfr - \bfrp|}
	\label{eq:E_H}
\end{equation}
is the classical Hartree energy. The XC energy $E_\text{xc}$ is often separated into the exchange and correlation
parts
\begin{equation}
	E_\mathrm{xc}[n] = E_\mathrm{x}[n]+E_\mathrm{c}[n]\, .
\end{equation}
Among the four terms in Eq.~\ref{eq:E_KS-DFT}, only $E_\text{ext}[n]$ and $E_\text{H}[n]$ are explicit functionals of $n(\bfr)$. 
The noninteracting kinetic energy $T_\text{s}$ is treated exactly in KS-DFT \index{KS-DFT} in terms of the single-particle KS orbitals $\psi_i(\bfr)$, 
which themselves are a functional of $n(\bfr)$.  All the many-body complexity is encoded in the unknown, but relatively small term $E_\mathrm{xc}[n]$.
Moreover, we often define the \emph{exact} exchange functional as the Hartree-Fock exchange, 
utilizing again the single-particle KS orbitals $\psi_{m\sigma}(\bfr)$
\begin{equation}
	\label{eq:E_HF_x}
	E_\mathrm{x}^\mathrm{EX}[n] = -\frac{1}{2}\sum_{mn,\sigma}^{occ}\langle \psi_{m\sigma}\psi_{n\sigma}|\psi_{n\sigma}|\psi_{m\sigma}\rangle.
\end{equation}

The notation of $\langle \psi_{p\sigma}\psi_{q\sigma'} |\psi_{r\sigma}\psi_{s\sigma'} \rangle$ 
is equivalent to $(\psi_{p\sigma}\psi_{r\sigma}|\psi_{q\sigma'}\psi_{s\sigma'})$, representing 
the two-electron Coulomb integrals,

\begin{equation}
	 \langle \psi_{p\sigma}\psi_{q\sigma'}|\psi_{r\sigma}\psi_{s\sigma'}\rangle 
= \iint d\bfr d\bfrp 
\frac{\psi_{p\sigma}^\ast(\bfr)\psi_{r\sigma}(\bfr)\psi_{q\sigma'}^\ast(\bfrp)\psi_{s\sigma'}(\bfrp)}{|\bfr-\bfrp|} =
	 ( \psi_{p\sigma}\psi_{r\sigma}|\psi_{q\sigma'}\psi_{s\sigma'}),
      \label{eq:2eri}			  
\end{equation}
For completeness, a relevant notation can be defined, which  will be used to express the second-order perturbation theory (PT2) in the following
section:
\begin{equation}
	 \langle \psi_{p\sigma}\psi_{q\sigma'}||\psi_{r\sigma}\psi_{s\sigma'}\rangle = \langle \psi_{p\sigma}\psi_{q\sigma'}|\psi_{r\sigma}\psi_{s\sigma'}\rangle- \delta_{\sigma\sigma'}\langle \psi_{p\sigma}\psi_{q\sigma}|\psi_{s\sigma}\psi_{r\sigma}\rangle.
      \label{eq:ce}
\end{equation}
Apparently, $E_{x}^\mathrm{EX}$ is also an implicit but not explicit functional of $n(\bfr)$. 
Here and below, we adopt the convention that $m,n$ refer to occupied KS orbitals, $a,b$ to virtual (unoccupied) orbitals, and $p,q,r,s$ to general ones. Furthermore, $\sigma$ and $\sigma'$ are the spin indices. Throughout this review, only the spin collinear case is considered in
the presented formulae.

In order to determine the single-particle KS orbitals, $\psi_{p\sigma}(\bfr)$, the so-call KS non-interacting system is introduced with 
the corresponding Hamiltonian written as
\begin{equation}
	\label{eq:H_KS}
	\begin{split}
		\hat{H}_0 = &\sum_{i}^{N} \left(- \frac{\nabla^2_i}{2} + v_\mathrm{KS}(\bfr_i) \right) ,
    \end{split}
\end{equation}
where the effective KS potential is defined as 
\begin{equation}
	\label{eq:V_ks}
	v_\mathrm{KS}(\bfr) = v^\mathrm{ext}(\bfr) + v_\mathrm{H}(\bfr) +v_{xc}(\bfr).
\end{equation}
Here the external potential $v^\mathrm{ext}(\bfr)$, the Hartree potential
\begin{equation}
	\label{eq:V_H}
   v_\mathrm{H}(\bfr) =  \frac{\delta E_\mathrm{H}[n]}{\delta n(\bfr)} =\int d\bfrp~ \frac{n(\bfrp)}{|\bfr - \bfrp|}\, ,
\end{equation}
and the XC potential, 
\begin{equation}
    \label{eq:V_xc}
	v_\mathrm{xc}(\bfr) =  \frac{\delta E_\mathrm{xc}[n]}{\delta n(\bfr)}=\frac{\delta (E_\mathrm{x}[n]+E_\mathrm{c}[n])}{\delta n(\bfr)}
	   = v_\mathrm{x}(\bfr) + v_\mathrm{c}(\bfr)\, ,
\end{equation}
if being exact, ensure that the ground-state wave function $\Phi_0$ of this KS non-interacting system, 
i.e. a Slater determinant composed by the aforementioned occupied single-particle KS orbitals 
$\psi_i(\bfr)$, 
reproduces the density $n(\bfr)$ of the true physical system $n(\bfr) = \langle \Psi_0 | \hat{n}(\bfr)|\Psi_0 \rangle$,
with $|\Psi_0\rangle$ being the interacting ground state. 
It is worth to note that the ground-state total energy of the KS non-interacting system $E_0$ 
\begin{equation}
	E_0[n] = \sum_{m,\sigma}^{occ}\epsilon_{m\sigma}=T_\text{s}[\psi] + \int v_\mathrm{KS}(\bfr) d\bfr = T_\text{s}[\psi] + E_\text{ext}[n] + 2E_\text{H}[n] + V_\text{xc}[n]\, ,
    \label{eq:E_KS-DFT_0}
\end{equation}
is not the same as that of the interacting one (Eq.~\ref{eq:E_KS-DFT}).
In Eq.~(\ref{eq:E_KS-DFT_0}), $\epsilon_{m\sigma}$'s are KS orbital energies and
\begin{equation}
	V_\text{xc}[n]=\int n(\bfr)v_\mathrm{xc}(\bfr)d\bfr\, ,
 \label{eq:V_xc_energy}
\end{equation}
is the expectation value of the XC potential (Eq.~\ref{eq:V_xc}).

A central task of the DFT community is to develop accurate and tractable approximations to $E_\mathrm{xc}[n]$. The success of KS-DFT lies in
the fact that usefully accurate approximations can be found, under which DFT calculations can be done at affordable cost. 
Widely used approximations to $E_\mathrm{xc}[n]$ include the aforementioned LDA \cite{Kohn/Sham:1965}, 
GGAs \cite{Becke:1988,Lee/Yang/Parr:1992,Perdew/Burke/Ernzerhof:1996}, 
meta-GGAs \cite{Tao/etal:2003,zhao_new_2006,Sun/Ruzsinszky/Perdew:2015,mn15-l:2016}, and hybrid functional approximations
\cite{Becke:1993,Heyd/Scuseria/Ernzerhof:2003,zhao_new_2006,mn15:2016,head-gordon:2017}.  
As discussed above, these approximations belong to the first four rungs of the Jacob's ladder \index{Jacob's ladder} \cite{Perdew/Schmidt:2001}.
Despite their enormous success, these existing approximate functionals suffer from many-electron self-interaction errors \cite{Perdew/etal:1982}, 
or delocalization errors \cite{Cohen/Mori-Sanchez/Yang:2008}, due to which the localized electronic states tend to delocalize over the system, 
leading to several severe consequences \cite{Cohen/Mori-Sanchez/Yang:2008}. Furthermore, the non-local correlation effects between widely separated subsystems
are not captured within these approximations by constructions. 
These intrinsic deficiencies limit the possible accuracy that can be achieved in practical 
calculations. Quantitative, and sometimes qualitative
 failures have been documented in a number of situations, among which the most prominent are van der Waals (vdW) \index{van der Waals (vdW)} 
bonded systems \cite{Kristyan/Pulay:1994}, materials with strong correlations \cite{Imada/Fujimori/Tokura:1998}, certain surface adsorption problems 
\cite{Feibelman:2001}, and chemical reaction barrier heights \cite{Zhang/Yang:1998}.
To deal with long-range vdW interactions, e.g., lower-rung functionals
often have to be complemented by semi-empirical vdW correction terms in various forms
\cite{Grimme:2006,Grimme/etal:2010,Becke/Johnson:2007,Tkatchenko/Scheffler:2009,Tkatchenko/etal:2012}.

Fifth-rung DFAs, including DHAs and RPA, can be derived within the framework of the so-called adiabatic-connection fluctuation-dissipation theorem (ACFDT) 
\index{adiabatic-connection fluctuation-dissipation theorem (ACFDT)}, which offers a powerful mathematical device to construct the exact 
XC energy functional via the density response functions \index{density response function} 
of a series of fictitious systems with scaled interactions, connecting the KS system and the true physical system. 
In this formulation, an approximation to the
density response function \index{density response function} translates into a corresponding approximation to the XC energy functional. 

\subsection{Adiabatic connection approach to DFT }
Within the adiabatic connection (AC) \index{the adiabatic connection (AC)} approach to DFT \cite{Langreth/Perdew:1977,Gunnarsson/Lundqvist:1976}, 
one considers a continuous set of fictitious Hamiltonian's,
   \begin{equation}
	   \label{eq:H_lambda}
	   \begin{split}
		   \hat{H}(\lambda) &= \hat{T}+\hat{V}_{\lambda}+\lambda\hat{V}_{ee} 
	    =-\sum_i^{N} \frac{\nabla^2_i}{2} + \sum_i^{N} v^\mathrm{aux}_\lambda(\hat{\bfr}_i) +
	        \frac{\lambda}{2} \sum_{i\ne j}^{N} \frac{1}{|\hat{\bfr}_i - \hat{\bfr}_j|} \, ,
	   \end{split}
   \end{equation}
which connects $\hat{H}_0$ at $\lambda=0$ with $v^\mathrm{aux}_{\lambda=0}=v_\mathrm{KS}=v^\mathrm{ext} + v_\mathrm{H}+v_\mathrm{xc}$ 
(Eq.~\ref{eq:V_ks}) and $\hat{H}$ at $\lambda=1$ with $v^\mathrm{aux}_{\lambda=1}=v^\mathrm{ext}$. 
The Hamiltonian $\hat{H}(\lambda)$ for $0<\lambda<1$ 
describes a collection of particles moving under the auxiliary external potential $v^\mathrm{aux}_\lambda(\bfr)$ and interacting 
with a scaled Coulomb interaction $\lambda/|\bfr-\bfrp|$.  The auxiliary 
potential $v^\mathrm{aux}_\lambda(\bfr)$ ($0<\lambda<1$) is chosen such that the density of $\lambda$-scaled systems is kept at the physical density,
i.e., $n_\lambda(\bfr)=n(\bfr)$, along the AC path. By utilizing the coordinate scaling technique, it can be shown that
\begin{equation}
	v^\mathrm{aux}_{\lambda}(\bfr) = v_\mathrm{KS}(\bfr) -\lambda \left[v_\mathrm{H}(\bfr)+ v_\mathrm{x}(\bfr) 
	+ \lambda\frac{\delta E_c[n_{1/\lambda}]}{\delta n(\bfr)}\right].
\end{equation}
Here $E_c[n_{1/\lambda}]$ is the correlation energy with respect to the scaled density (see Ref. \cite{Goerling/Levy:1993,levy:1985} 
for more detailed definition and derivation).
$E_c[n_{1/\lambda}]=E_c[n] $ when $\lambda=1$; while in the high density limit with $\lambda=0$, $E_c[n_{1/\lambda}]=0$.
For the fictitious system of $\hat{H}(\lambda)$ with fixed density $n(\bfr)$, the ground-state wavefunction can be denoted as $|\Psi_\lambda^{n}\rangle$, 
         \begin{equation}
			 \hat{H}(\lambda)|\Psi_\lambda^{n}\rangle = E_\lambda |\Psi_\lambda^{n}\rangle
         \end{equation}
with the normalization condition of $\langle \Psi_\lambda^{n} | \Psi_\lambda^{n}\rangle=1$. 
Here $E_{\lambda}$ is the ground-state energy along the AC path. 
$E_{0} = E_{\mathrm{KS}}$ is nothing but the ground-state energy of non-interacting KS system (Eq.~\ref{eq:E_KS-DFT_0})
, while $E_{1} = E$ is the ground-state energy of the fully interacting system (Eq.~\ref{eq:E_KS-DFT}).

The Hellman-Feynman theorem implies that
\begin{equation} \label{eq:HFT}
	\begin{split}
      \frac{dE_\lambda}{d\lambda} & = 
	  \left\langle \Psi_\lambda^n \left| \frac{\partial \hat{H}(\lambda)}{\partial \lambda} \right| \Psi_\lambda^{n} \right \rangle  
	  = \left\langle \Psi_\lambda^{n} \left|\frac{\partial \hat{V}_{\lambda}}{\partial \lambda}\right| \Psi_\lambda^{n} \right\rangle
	  +\left\langle \Psi_\lambda^{n} \left| \hat{V}_{ee}  \right| \Psi_\lambda^{n} \right\rangle \\ 
	  & = \left \langle \Psi_\lambda^{n} \left|\sum_{i=1}^N \frac{\partial v^\mathrm{aux}_\lambda(\hat{\bfr}_i)}{\partial \lambda} +  
	  \frac{1}{2} \sum_{i\ne j}^{N} \frac{1}{|\hat{\bfr}_i - \hat{\bfr}_j|}  \right| \Psi_\lambda^{n} \right\rangle \\ 
      & = \int d\bfr n(\bfr) \frac{\partial v^\mathrm{aux}_\lambda(\bfr)}{\partial \lambda} +  
	  \frac{1}{2} \iint d\bfr d\bfrp \frac{\langle \Psi_\lambda^{n} \left|\hat{n}(\bfr)\hat{n}(\bfrp)\right|\Psi_\lambda^{n} \rangle
                -n(\bfr)\delta(\bfr-\bfrp)} {|\bfr - \bfrp|} 
	\end{split}
\end{equation}
Here, we have used the expression for the density operator of the $N$-electron systems
   \begin{equation}
	   \hat{n}(\bfr) = \sum_{i=1}^N \delta(\bfr - \hat{\bfr}_i) \, ,
	   \label{eq:density_operator}
   \end{equation}
and the condition $\langle \Psi_\lambda |\hat{n}(\bfr)| \Psi_\lambda\rangle=n(\bfr)$.
By integrating Eq.~\ref{eq:HFT} over the whole adiabatic connection path, the ground-state energy of the interacting
system can be obtained as 
   \begin{equation}
       \label{eq:E_int_AC}
	   \begin{split}
	   E  = & E_0  +  \int_0^1 \frac{dE_\lambda}{d\lambda} d\lambda  \\
          = & E_0  +  \int d\bfr~ n(\bfr) \left[ v^\mathrm{aux}_{\lambda=1}(\bfr) - v^\mathrm{aux}_{\lambda=0}(\bfr) \right] 
		+\int_0^1 \left\langle \Psi_\lambda^{n} \left|\hat{V}_{ee}  \right| \Psi_\lambda^{n} \right\rangle d\lambda\\
		  = & T_s[\psi]+ E_\mathrm{ext}[n] +\int_0^1 \left\langle \Psi_\lambda^{n} \left|\hat{V}_{ee}  \right| \Psi_\lambda^{n} \right\rangle d\lambda\\
		  = & T_s[\psi]+ E_\mathrm{ext}[n] + E_\mathrm{H}[n] \\
		    &+\int_0^1 d\lambda~ \frac{1}{2} \iint d\bfr d\bfrp~ 
		    \frac{\langle \Psi_\lambda^{n} \left|\delta \hat{n}(\bfr)\delta \hat{n}(\bfrp)\right|\Psi_\lambda^{n} \rangle
            -n(\bfr)\delta(\bfr-\bfrp)} {|\bfr - \bfrp|} \, ,
	 \end{split}
   \end{equation}
where we have introduced the fluctuation of the density operator $\delta \hat{n}(\bfr) = \hat{n}(\bfr) - n(\bfr)$, and used the fact that 
$\langle \Psi_\lambda | \delta \hat{n}(\bfr) | \Psi_\lambda \rangle =0$ and the definitions of $E_{\mathrm{H}}$,  $E_0[n]$,
and $V_{\mathrm{xc}}$
(Eqs.~\ref{eq:E_H}, \ref{eq:E_KS-DFT_0}, and \ref{eq:V_xc_energy}). 
Comparing Eq.~\ref{eq:E_KS-DFT} to Eq.~\ref{eq:E_int_AC} yields the XC energy in terms of the coupling-constant integration
\begin{equation}
	\label{eq:Exc_AC}
	E_{\mathrm{xc}}[n]= \int_0^1 W_{\lambda}[n]d\lambda,
\end{equation}
with the XC potential along the AC path $W_{\lambda}$ defined as
\begin{equation}
	\label{eq:W_lambda}
	\begin{split}
		W_{\lambda}[n]
               = & \left\langle \Psi_\lambda^{n} \left|\hat{V}_{ee} \right| \Psi_\lambda^{n} \right\rangle - E_{\mathrm{H}}[n]\\
			   = &\frac{1}{2} \iint d\bfr d\bfrp~ 
				\frac{\langle \Psi_\lambda^{n} \left|\delta \hat{n}(\bfr)\delta \hat{n}(\bfrp)\right|\Psi_\lambda^{n} \rangle
				-n(\bfr)\delta(\bfr-\bfrp)} {|\bfr - \bfrp|}.\\
			\end{split}
\end{equation}
It is easy to see that the XC potential at the starting point with $\lambda = 0$ and $\Psi_0^{n}=\Phi_0$ is nothing but the exact exchange energy
(Eq.~\ref{eq:E_HF_x}) in the KS framework,
\begin{equation}
	\label{eq:W_lambda0}
	\begin{split}
		W_{0}[n] = & \left\langle \Phi_{0} \left|\hat{V}_{ee} \right| \Psi_{0} \right\rangle - E_{\mathrm{H}}[n] = E_\mathrm{x}^\mathrm{EX}[n].\\
	\end{split}
\end{equation}
If $W_{\lambda}$ were known for the whole AC path, we would be able to approach the exact XC functional via the above coupling-constant
integration (Eq. \ref{eq:Exc_AC}). 
From this perspective, better DFAs can be designed by making a better description of the integrands $W_{\lambda}$ along the AC path.

\subsection{G\"orling-Levy perturbation theory and XYG3-type DHAs}
\label{sec:GL2}
Let us start by reformulating the Hamiltonian $\hat{H}_{\lambda}$ (Eq.~\ref{eq:H_lambda}) as
\begin{equation}
	\hat{H}_{\lambda} = \hat{H}_0 + \lambda\hat{H}',
\end{equation}
where $\hat{H}_0$ is the Hamiltonian of the KS non-interacting system (Eq.~\ref{eq:H_KS}) and the perturbation 
$\hat{H}'$ is therefore 
\begin{equation}
	\label{eq:H'}
	\begin{split}
	   \hat{H}' = &\hat{V}_{ee}+\frac{1}{\lambda}\sum_{i}^{N}[v_{\lambda}^\mathrm{aux}(\bfr_i)-v_\mathrm{KS}(\bfr_i)]\\
	   =&\hat{V}_{ee}-\sum_{i}^{N}\left[v_\mathrm{H}(\bfr_i)+v_\mathrm{x}(\bfr_i)
	   +\lambda\frac{\delta E_c[n_{1/\lambda}]}{\delta n(\bfr_i)}\right].
    \end{split}
\end{equation}
Eq.~\ref{eq:H'} suggests that the perturbation along the AC path is almost linear, except for the correlation part which evolves nonlinearly
with respect to the coupling-constant parameter $\lambda$.
As a consequence, the perturbative $\hat{H}'$ for $\lambda\rightarrow 0$ is written down as
\begin{equation}
	\Delta=\lim_{\lambda\rightarrow 0}\hat{H}' = \hat{V}_{ee}-\sum_i^N[v_\mathrm{H}(\bfr_i)+v_x(\bfr_i)].
\end{equation}
According to the standard perturbation theory, the ground-state energy of $\hat{H}_{\lambda}$ can be expanded perturbatively based on the 
ground-state energy of the KS non-interacting system $E_{0}$ (Eq. \ref{eq:E_KS-DFT_0})
\begin{equation}
	\label{eq:GLPT}
	E_{\lambda}=E_0 + \lambda E^{(1)} + \lambda^2 E^{(2)} + \lambda^3 E^{(3)} + O(\lambda^4),
\end{equation}
where $E^{(k)}$ refers to the $k$th-order energy correction to $E_0$. 
In particular, the first-order energy correction for $\lambda=0$ is 
\begin{equation}
	\label{eq:GL1}
	\left.E^{(1)}\right|_{\lambda=0} = \left.\frac{\partial E_{\lambda}}{\partial \lambda}\right|_{\lambda=0}  
	= \left\langle \Psi_0^{n} \left|\Delta \right| \Psi_0^{n} \right\rangle 
	= E_{x}^{\mathrm{EX}} - E_\mathrm{H} - V_x,\\
\end{equation}
and the second-order energy correction for $\lambda=0$ is
\begin{equation}
	\label{eq:GL2}
	\begin{split}
		\left.E^{(2)}\right|_{\lambda=0}=\frac{1}{2}\left.\frac{\partial^2 E_{\lambda}}{\partial \lambda^2}\right|_{\lambda=0}
				=E_{c}^{\mathrm{GL2}}=E_{c}^\mathrm{GL-SE}+E_{c}^\mathrm{PT2}\\
	\end{split}
\end{equation}
which is widely recognized as the G\"orling-Levy (GL) theory of the coupling-constant perturbation expansion to the second order \cite{Goerling/Levy:1993}. 
The first term of $E_c^\mathrm{GL2}$ is associated with the single excitation (SE) contribution
\begin{equation}
	\label{eq:GL-SE}
	E_c^\mathrm{GL-SE}=\sum_{m}^{occ}\sum_a^{vir}\sum_{\sigma}
\frac{\left|\left\langle\psi_{m\sigma}\left|v_x^\mathrm{HF}-v_x\right|\psi_{a\sigma}\right\rangle\right|^2}{\varepsilon_{m\sigma}-\varepsilon_{a\sigma}},
\end{equation}
where $v_x^\mathrm{HF}$ is the Hartree-Fock non-local exchange potential, which is the derivative of $E_{x}^\mathrm{EX}$ against the KS orbitals,
while $v_x$ is the KS local exchange potential defined in Eq.~\ref{eq:V_xc}. 
The second part of $E_c^\mathrm{GL2}$ counts the contributions from double excitations, namely $E_c^\mathrm{PT2}$, 
\begin{equation}
	\label{eq:PT2}
	E_{c}^\mathrm{PT2}=
		\frac{1}{4}\sum_{m,n}^{occ}\sum_{a,b}^{vir}\sum_{\sigma,\sigma'}\frac{\left|\left\langle\psi_{m\sigma}\psi_{n\sigma'}||\psi_{a\sigma}\psi_{b\sigma'}\right\rangle\right|^2}
		{\varepsilon_{m\sigma}+\varepsilon_{n\sigma'}-\varepsilon_{a\sigma}-\varepsilon_{b\sigma'}},
\end{equation}
where  $\langle\psi_{m\sigma}\psi_{n\sigma'}||\psi_{a\sigma}\psi_{b\sigma'}\rangle$  is defined in Eq.~\ref{eq:ce}.
$E_c^\mathrm{PT2}$ shares the same formula as the second-order M{\o}ller-Plesset perturbation theory (MP2). 

As discussed above, the initial integrand along the AC path is given by Eq.~\ref{eq:W_lambda0}, which is simply 
the Hartree-Fock like exact-exchange energy (Eq.~\ref{eq:E_HF_x}).
Moreover, the second-order derivative against $\lambda$ based on Eq.~\ref{eq:HFT} together with Eq.~\ref{eq:GL2}
defines the initial slope of $W_{\lambda}$
\begin{equation}
	\begin{split}
		W_0'[n] = &\left. \frac{\partial W_{\lambda}[n]}{\partial \lambda}\right|_{\lambda=0} = 2E^{\mathrm{GL2}}.
	\end{split}
\end{equation}

\begin{figure}[t]
 \centering
    \includegraphics[scale=1.2]{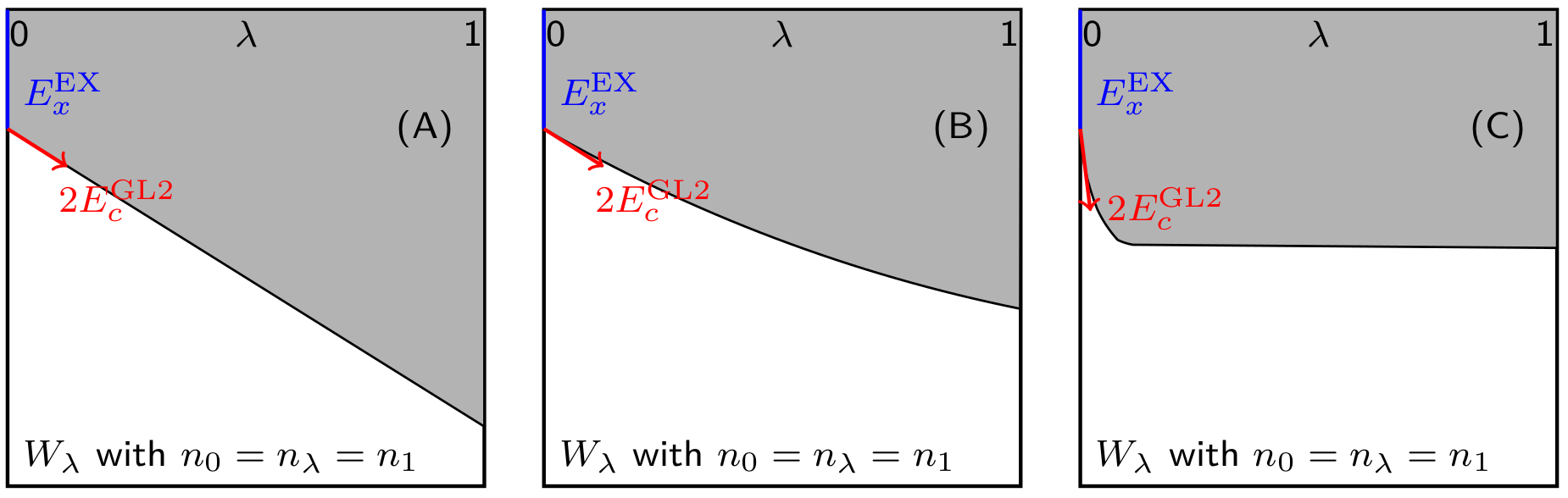}
	\caption{Illustration of adiabatic-connection approach in KS-DFT: (A) a linear AC path; (B) a non-linear AC path; (C) an AC path
		with a large initial slope.}
 \label{fig:AC_path}
\end{figure}

If the XC potential $W_{\lambda}$ along the AC path is linear (see Figure \ref{fig:AC_path}A), it is precisely determined by 
$W_{0}$ and $W_{0}'$
\begin{equation}
	W_\lambda^{\mathrm{linearAC}} = W_{0}+W_{0}'*\lambda = E_{x}^{\mathrm{EX}}+2E_{c}^\mathrm{GL2}*\lambda.
\end{equation}
As a result, the corresponding XC functional $E_{xc}^{\mathrm{linearAC}}$ can be obtained by applying Eq.~\ref{eq:Exc_AC},
\begin{equation}
	\label{eq:linearAC}
	E_{xc}^{\mathrm{linearAC}} = E_{x}^{\mathrm{EX}}+E_{c}^\mathrm{GL2}.
\end{equation}
Eq.~\ref{eq:linearAC} looks similar to the standard MP2 method, which is composed of the HF exchange and the MP2 correlation. 
However, it has to be emphasized that the MP2 method is the lowest-level correlated
wavefunction method for many-electron systems, while Eq.~\ref{eq:linearAC} is the \emph{exact (very accurate) exchange-correlation functional} 
for any systems with linear (quasi-linear) $\lambda$-dependence. The key difference here is that, 
the Hartree-Fock orbitals are used for the standard MP2 calculations, while the KS 
non-interacting orbitals, which deliver the exact density of the interacting system, are employed for the evaluation of 
HF-like exchange and GL2 correlation in Eq.~\ref{eq:linearAC}.
The linear AC functional in the context of coupling-constant expansion indicates that, for the development of fifth-rung DFAs,
the KS unoccupied orbitals can be introduced in the form of second-order perturbation theory. 
Equally important is the quality of input density and orbitals, which ensures the double of $E_{c}^\mathrm{GL2}$ properly represents the initial
slope of the XC potential along the AC path. 

In this consideration, XYG3-type DHAs were proposed to go beyond the linear model by introducing several empirical parameters to extrapolate
the second-order perturbation to infinite order (see Figure \ref{fig:AC_path}B). For example, the XYG3-type DHAs utilizing the 
B3LYP orbitals (xDH@B3LYP)\cite{zhang:2021A} has the form of 
\begin{equation}
	\label{eq:xdh@b3lyp}
	\begin{split}
	E_{xc}^\mathrm{xDH@B3LYP} = &a_1 E_{x}^\mathrm{EX}+a_2 E_x^\mathrm{S}+a_3 E_x^\mathrm{B88}+a_4 E_c^\mathrm{VWN}\\
	&+a_5E_c^\mathrm{LYP} +a_6E_c^\mathrm{osPT2} +a_7E_c^\mathrm{ssPT2}.
    \end{split}
\end{equation}
where $E_x^\mathrm{S}$ and $E_x^\mathrm{B88}$ are the exchange contribution from the Slater-type LDA and 
the Becke88 GGA. $E_c^\mathrm{VWN}$ is the local Vosko-Wilk-Nusair correlation, and $E_c^\mathrm{LYP}$ is the Lee-Yang-Parr 
correlation approximation. Here, $E_c^\mathrm{osPT2}$ and $E_c^\mathrm{ssPT2}$ are the opposite-spin and same-spin part of 
PT2 (Eq.~\ref{eq:PT2}), respectively.
\begin{equation}
	E_c^\mathrm{osPT2} +E_c^\mathrm{ssPT2} = E_c^\mathrm{PT2}. 
\end{equation}
It is easy to find that in xDH@B3LYP, the single-excitation contribution ($E_c^\mathrm{SE}$) is not taken into account. It was argued that
$E_c^\mathrm{SE}$ can be absorbed into the fitting parameters of xDH@B3LYP against the experimental data. Note that, the importance of $E_c^\mathrm{SE}$
was revisited by Ren et. al. \cite{Ren/etal:2011} for improving the RPA method in describing vdW-bonded molecules. This will be discussed in Section \ref{sec:beyond_rpa}.

Equation~\ref{eq:xdh@b3lyp} suggests that there are at most 7 empirical parameters in the general formula of xDH@B3LYP methods. In fact,
any xDH@B3LYP methods can be written down using this formula. The first approximation in this family, i.e.,\ XYG3 proposed in 2009 \cite{zhang:2009A}, 
contains only 3 parameters, indicating that four constraints are imposed during the parameter optimization.
Similarly, XYGJ-OS proposed in 2011 contains 4 parameters with 3 constraints \cite{zhang:2011A}. Recently, the accuracy limit of 
xDH@B3LYP methods was explored by fully optimizing all 7 parameters. The resulting XYG7 functional \cite{zhang:2021A} is among the top-class performers 
in the comprehensive benchmark for main-group chemistry,
which can provide a balanced description of both covalent and noncovalent interactions. Its accuracy is comparable to or even better than the very
expensive composite methods in wave function theory.

A key deficiency of PT2-based DHAs is in the description of strong correlation systems, where 
the degeneration of low-lying states could result in extremely large correlation contributions based on the perturbation treatment at any finite orders.
In the context of coupling-constant expansion along the AC path, it means that the initial slope $W_0'[n]=2E_{c}^\mathrm{GL2}$ becomes extremely large,
such that the initial slope itself is of little relevance to the exact XC energy, which is mainly determined by the XC
potential at the fully interacting system $W_{\lambda=1}[n]$ (see Figure \ref{fig:AC_path}C).

To address this problem, we should go beyond the finite-order perturbation treatment of the correlation effects. Among different kinds
of solutions already proposed, RPA is a promising candidate, as will be introduced in the following section. 
It offers a systematic way to renormalize different contributions in PT2, and thus
address the divergence problem of finite-order perturbation theory as well as DHAs for strong-correlation systems.

\subsection{RPA derived from the ACFDT framework}
\label{sec:ACFDT_RPA}

Let us return to the coupling-constant integration formula of the XC functional.
From the viewpoint of zero-temperature fluctuation-dissipation theorem \cite{Nozieres/Pines:1966},
the density-density correlation (\textit{fluctuations}) in Eqs.~\ref{eq:Exc_AC} and \ref{eq:W_lambda} 
can be related to the imaginary part of the density response function \index{density response function}
\index{density response function} (\textit{dissipation}) of the system 
\begin{equation}
 \langle \Psi_\lambda|\delta\hat{n}(\bfr)\delta\hat{n}(\bfrp)|\Psi_\lambda\rangle = -
  \frac{1}{2\pi}\int_{-\infty}^\infty d\omega \text{Im} \chi^\lambda (\bfr, \bfrp, i\omega)\, .
 \label{eq:FDT}
\end{equation}
Here the density response function $\chi^\lambda (\bfr, \bfrp, \omega)=\delta n_\lambda(\bfr,\omega)/\delta v^{ext}(\bfrp,\omega)$ describes the variation of 
the density of the partially interacting system at the spatial point $\bfr$, up to the linear order, due to a change of the local external potential at $\bfrp$.  
From Eqs.~\ref{eq:Exc_AC}, \ref{eq:W_lambda} and \ref{eq:FDT},  we arrive at the renowned ACFDT expression for the XC energy in DFT,
\begin{equation}
 \label{eq:EC-ACFD}
	\begin{split}
    E_\text{xc} =& \frac{1}{2} \int_0^1 d\lambda \iint d\bfr d\bfrp \frac{1}{|\bfr-\bfrp|}
            \left[ 
	-\frac{1}{2\pi}\int_{-\infty}^\infty d\omega \text{Im} \chi^\lambda (\bfr, \bfrp, \omega) - \delta(\bfr-\bfrp)n(\bfr) 
           \right] \\
       =& \frac{1}{2} \int_0^1 d\lambda \iint d\bfr d\bfrp  \frac{1}{|\bfr-\bfrp|}
        \left[  -\frac{1}{\pi}\int_0^\infty d\omega 
        \chi^\lambda (\bfr, \bfrp, i\omega) - \delta(\bfr-\bfrp)n(\bfr) \right]\,  . 
	\end{split}
\end{equation}
The fact that the above frequency integration can be performed along the imaginary frequency axis originates
from the pole structure of $\chi^\lambda (\bfr, \bfrp, \omega)$ and the fact that it becomes purely
real on the imaginary axis. Such a property simplifies the ground-state energy calculation within the ACFDT \index{ACFDT} framework considerably. 
The ACFDT expression in Eq.~\ref{eq:EC-ACFD} translates the problem of computing the XC energy to the computation of
the response functions of a continuous set of fictitious systems along the AC path, which in practice have to be approximated as well. 

Obviously, the exact density response function \index{density response function} $\chi^\lambda (\bfr, \bfrp, i\omega)$ is not known either. 
However, according to linear-response time-dependent DFT (TDDFT),
the interacting response function $\chi^\lambda$ for $\lambda>0$ is linked to the noninteracting response function $\chi^0$ via the 
Dyson-like equation
\begin{equation}
	\begin{split}
    \chi^{\lambda}(\bfr, \bfrp, i\omega) = &\chi^0(\bfr,\bfrp,i\omega) + \int d\bfr_1 \int d\bfr_2 \chi^0(\bfr,\bfr_1,i\omega)\\
	  &\times \left[\frac{\lambda}{|\bfr_1-\bfr_2|}+f_\mathrm{xc}^\lambda(\bfr_1,\bfr_2,i\omega) \right] \chi^{\lambda}(\bfr_2,\bfrp,\omega) \, ,
   \end{split}
 \label{eq:LR_TDDFT}
\end{equation}
where $f_\mathrm{xc}^\lambda$ is the so-called exchange-correlation kernel, which in time domain is given by
  \begin{equation}
	  f_\mathrm{xc}^\lambda(\bfr_1,\bfr_2,t-t') = \frac{\delta v_\mathrm{xc}^\lambda(\bfr_1,t)}{\delta n(\bfr_2,t')} = 
	  \frac{\delta^2 A^\lambda_\mathrm{xc}[n]}{\delta n(\bfr_1,t) \delta n(\bfr_2,t')}\, .
  \end{equation}
Here $A^\lambda_\mathrm{xc}$ is the exchange-correlation part of the action functional \cite{Onida/Reining/Rubio:2002}. 
     
Obviously, the $f_\mathrm{xc}$ kernel is a very complex quantity to deal with, and there are considerable ongoing efforts to 
find better approximations for it \cite{Goerling:2019}.  In this context, the \textit{random phase approximation} amounts to
simply neglecting the $f_\mathrm{xc}$ kernel, and the resultant interacting response function is termed as the RPA response function,
\begin{equation}
 \chi^{\lambda}_{\text{RPA}}(\bfr, \bfrp, i\omega) = \chi^0(\bfr,\bfrp,i\omega) + 
    \int d\bfr_1 d\bfr_2 \chi^0(\bfr,\bfr_1,i\omega)
   \frac{\lambda}{|\bfr-\bfrp|} \chi^{\lambda}_{\text{RPA}}(\bfr_2,\bfrp,\omega)\, . 
 \label{eq:RPA_response}
\end{equation}
In physical terms, the RPA here corresponds to linearized time-dependent Hartree approximation, by which the variation of the 
exchange-correlation potential due to an external perturbation is neglected.

In Eqs.~\ref{eq:LR_TDDFT} and \ref{eq:RPA_response}, $\chi^0(\bfr,\bfr_1,i\omega)$ is the independent-particle response function of the 
KS reference system ($\lambda=0$) and is known explicitly in terms of the single-particle KS orbitals $\psi_{p\sigma}(\bfr)$, 
orbital energies $\epsilon_{p\sigma}$, and occupation factors $f_{p\sigma}$,
\begin{equation}
\chi^0(\bfr,\bfrp,i\omega)  = \sum_{p,q,\sigma}\frac{(f_{p\sigma}-f_{q\sigma})\psi_{p\sigma}^\ast(\bfr)\psi_{q\sigma}(\bfr) \psi_{q\sigma}^\ast(\bfrp)
\psi_{p\sigma}(\bfrp)}{\epsilon_{p\sigma} - \epsilon_{q\sigma} -i\omega}\, .
\label{eq:indep_response}
\end{equation}
Based on Eqs.~\ref{eq:EC-ACFD} and \ref{eq:RPA_response}, the XC energy in RPA can be further decomposed 
into an exact exchange (EX) term and a RPA correlation term,
\begin{equation}
E_\mathrm{xc}^\mathrm{RPA}[n] = E_\mathrm{x}^\mathrm{EX}[\psi] + E_\mathrm{c}^\mathrm{RPA}[\epsilon,\psi],
	\label{eq:E_xc_RPA}
\end{equation}
where
\begin{equation}
	\begin{split}
 E_\text{x}^\text{EX}[\psi]
       =& -\frac{1}{2}\sum_{p,q,\sigma} f_{p\sigma} f_{q\sigma} \iint d\bfr d\bfrp \frac{\psi_{p\sigma}^\ast(\bfr) \psi_{q\sigma}(\bfr) 
         \psi_{q\sigma}^\ast(\bfrp)\psi_{p\sigma}(\bfrp)}{|\bfr-\bfrp|}\,  \label{eq:E_EX} 
	 \end{split}
\end{equation}
and
\begin{equation}
	\begin{split}
	E_\text{c}^\text{RPA}[\epsilon,\psi]  =& -\frac{1}{2\pi} \iint d\bfr d\bfrp  \frac{1}{|\bfr-\bfrp|}
       \int_0^\infty d\omega \left[ \int_0^1 d\lambda \chi^{\lambda}_{\text{RPA}}(\bfr, \bfrp, i\omega)
             -\chi_{0}(\bfr, \bfrp, i\omega) \right] \\
   =& \frac{1}{2\pi}\int_0^\infty d\omega \text{Tr} \left[ \text{ln}(1-\chi^0 (i\omega) v) + 
      \chi^0 (i\omega) v \right],
  \end{split}
 \label{eq:Ec_RPA}
\end{equation}
with $v(\bfr,\bfrp) = 1/|\bfr-\bfrp|$. Equation~\ref{eq:E_EX} defines the exact-exchange energy (Eq.~\ref{eq:E_HF_x}) within the ACFDT context, 
and is extended here to fractional occupation numbers and to be evaluated with KS orbitals.
Furthermore, in the second line of Eq.~\ref{eq:Ec_RPA}, for brevity the following convention 
 \begin{equation}
  \text{Tr}\left[fg\right] = \iint d\bfr d\bfrp f(\bfr,\bfrp)g(\bfrp,\bfr)
  \label{eq:trace}
 \end{equation}
has been adopted.

So far, we have described how the RPA method is defined as an approximate XC energy functional in the KS-DFT context, 
within the ACFDT \index{ACFDT} framework. As mentioned above, RPA can also be derived from the perspective of the coupled
cluster theory \index{coupled cluster theory} and the Green-function based  many-body perturbation theory\index{many-body perturbation theory}.
For instance, to link with the ring coupled cluster doubles (rCCD) theory, the RPA correlation energy can be obtained as,
         \begin{equation}
		    E_\text{c}^\text{RPA} = \frac{1}{2}\text{Tr}\left(BT^\text{rCCD}\right) = 
		       \frac{1}{2}\sum_{mn}^{occ}\sum_{ab}^{vir}\sum_{\sigma,\sigma'}B_{ma\sigma,nb\sigma'}T_{nb\sigma',ma\sigma}^\text{rCCD}\, .
              \label{eq:Ec_rpa_rccd}			  
	 \end{equation}
where $T_{nb\sigma',ma\sigma}^\text{rCCD}$ is the rCCD amplitude, to be determined by solving the so-called the Riccati equation 
\cite{Scuseria/Henderson/Sorensen:2008}, and $B_{ma\sigma,nb\sigma'}=\langle \psi_{m\sigma}\psi_{n\sigma'} | \psi_{a\sigma}\psi_{b\sigma'} \rangle$.
Due to the limited space, here we will not elaborate on the alternative formulations of RPA any further, and interested readers are referred to 
Ref.~\cite{Ren/etal:2012b} and the original references \cite{Scuseria/Henderson/Sorensen:2008,Dahlen/Leeuwen/Barth:2006} for more details.

Finally, combining Eqs.~\ref{eq:E_KS-DFT} and \ref{eq:E_xc_RPA}, one obtains the total
ground-state energy of the RPA method, 
\begin{align}
 E[n,\psi] &= T_\mathrm{s}[\psi] + E_\mathrm{ext}[n] + 
 E_\mathrm{H}[n] + E_\mathrm{x}^\mathrm{EX}[\psi]  +  E_\mathrm{c}^\mathrm{RPA}[\epsilon,\psi]  \nonumber \\
 & = \langle \Psi_{\lambda=0}^n|\hat{H}|\psi_{\lambda=0}^n\rangle + E_\mathrm{c}^\mathrm{RPA}[\epsilon,\psi]
 \label{eq:RPA_E_total}
\end{align}
where the first four terms can be grouped together, yielding an energy expression that
corresponds to the Hartree-Fock energy evaluated with KS orbitals. This is simply given by
the expectation value of the full interacting Hamiltonian within the KS Slater determinant.

In practice, RPA calculations are usually done perturbatively on top of a preceding calculation based on
lower-rung DFAs. That is, the single-particle orbitals $\{\psi_{p\sigma}\}$ and orbital energies
$\{\epsilon_{p\sigma}\}$ employed in Eq.~\ref{eq:RPA_E_total} are generated using lower-rung 
approximations such as LDA, GGAs, or meta-GGAs. Thus, practical RPA calculations are often denoted as ``RPA@DFA'', where ``DFA''
refers to the actual functional used in the preceding reference-state calculation. 

\subsection{\label{sec:beyond_rpa}Beyond-RPA computational schemes}

Despite its appealing features, RPA does not go without shortcomings. The conventional wisdom is that RPA describes the long-range 
correlation very well, whereas it is not adequate for short-range correlations. This issue has been well known for homogeneous electron 
gas. For real materials, RPA was found to underestimate the cohesive energies of both molecules \cite{Furche:2001,Ruzsinszky/etal:2010} 
and solids \cite{Harl/Schimka/Kresse:2010}. This stimulated much research interest and several beyond-RPA schemes have been developed to
fix this deficiency. Among these, three approaches are particularly noteworthy: 
1) Hybridizing RPA with the lower-rung DFAs in the empirical double hybrid framework; 
2) improving RPA by restoring the contribution of the $f_\mathrm{xc}$ kernel \index{$f_\mathrm{xc}$ kernel} within the ACFDT \index{ACFDT} formalism; 
and 3) improving RPA from the perspective of diagrammatic many-body perturbation theory.
The first empirical approach gains increasing attention in quantum chemistry. The proposed RPA-based DHAs includes 
(a) dRPA75 \cite{mezei/csonka/ruzsinszky/kallay:2015} and PWRB95 \cite{grimme/steinmetz:2016} where the direct RPA correlation is used to replace
the PT2 term in the double hybrid formula; (b) SCS-dRPA75 \cite{mezei/csonka/ruzsinszky/kallay:2017} and scsRPA \cite{Zhang/Xu:2019} that utilize
different kinds of spin-component scaled RPA; and (c) SCS-dRPA75rs \cite{Mezei/Kallay:2019} with range-separated RPA.  
Successful beyond-RPA schemes of the second type include the truncated adiabatic LDA kernel correction 
\cite{Olsen/Thygesen:2012}, and the more systematic construction of the XC kernel with respect to a series expansion of the coupling 
constant $\lambda$ within the ACFDT framework \cite{Erhard/etal:2016,Goerling:2019}. 

\begin{figure}[t]
 \centering
	\begin{picture}(400,220)(0,0)
	\put(60,0){\includegraphics[width=0.6\textwidth]{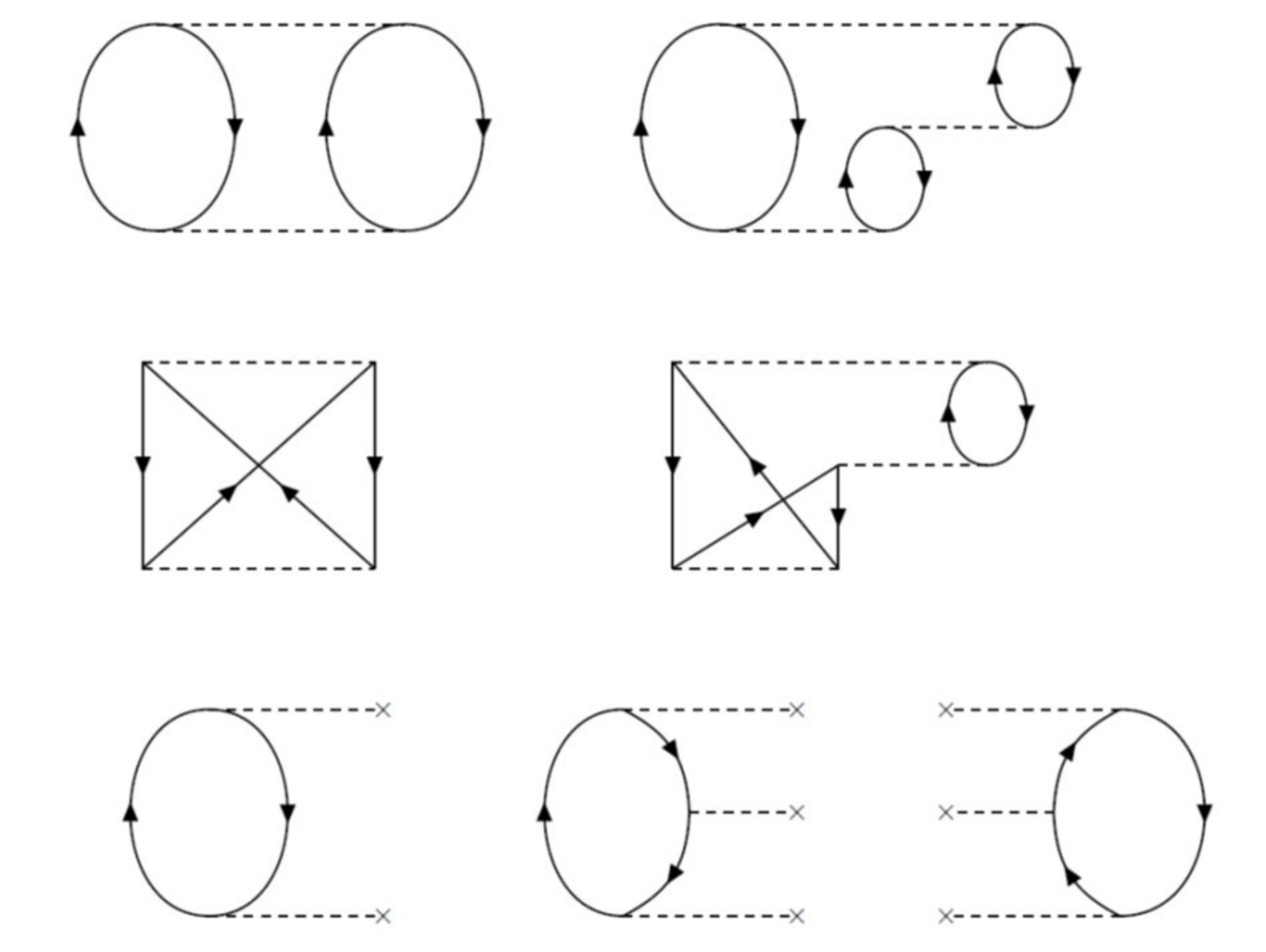}}
        \put(178,175){+}
        \put(300,175){+ $\cdots$}
        \put(167,105){+}
        \put(300,105){+ $\cdots$}
        \put(155,27){+}
        \put(244,27){+}
        \put(335,27){+ $\cdots$}
	\end{picture}
	\caption{Goldstone diagrams for rPT2.  Diagrams in the three row corresponds to RPA, SOSEX, and rSE \index{rSE} respectively.  
	 Dashed line ending with a cross in the third row denotes the matrix element
	    $\Delta v_{pq} = \langle \psi_p| \hat{v}^\text{HF} - \hat{v}^\text{MF} |\psi_q\rangle$. The rules
	        to evaluate Goldstone diagrams can be found in \refcite{Szabo/Ostlund:1989}.}
 \label{fig:rPT2_diagram}
\end{figure}
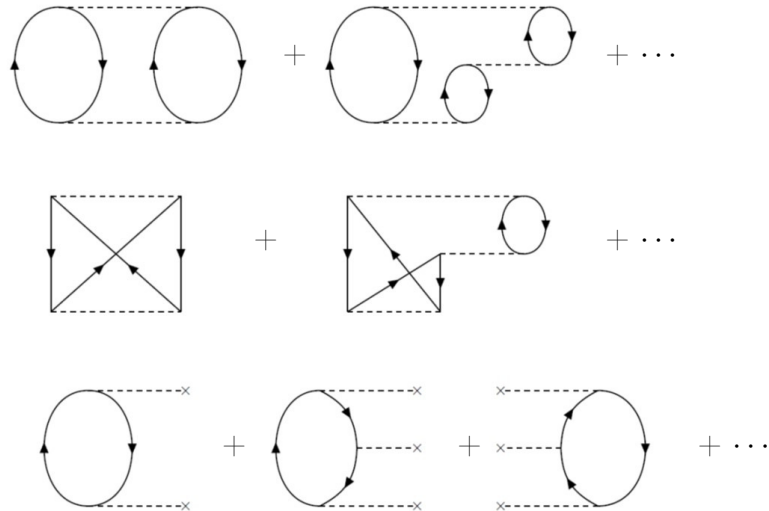

The diagrammatic many-body expansion approach to correct RPA, on which we shall concentrate here, exploits the fact that RPA, in contrast with other
density functional approximations (DFAs), has a clear diagrammatic representation -- an infinite summation of the ring diagrams. The question arises if
there is a simple and systematic way to incorporate the missing diagrams to arrive at an improved theory. The fermionic nature of electrons requires the
many-electron wavefunction to be antisymmetric, and the diagrammatic representation of the correlation energy contains graphs describing both 
\textit{direct} processes and \textit{exchange} processes. 
For example, in the M{\o}ller-Plesset perturbation theory, both types of processes are included at each order, which ensures 
the theory to be one-electron ``self-correlation free'' -- the correlation energy for an one-electron system being zero. The standard
PT2 correlation (Eq.~\ref{eq:PT2})
contains one \textit{direct} term and one \textit{exchange} term, represented respectively by the leading diagram in the first two rows of 
Fig.~\ref{fig:rPT2_diagram}. These two terms cancel each other for
one-electron systems. RPA is represented by the ring diagrams\index{ring diagrams} summed up to infinite order, as represented by the first-row diagrams in  
Fig.~\ref{fig:rPT2_diagram}, where only ``direct'' processes are accounted for \footnote{The RPA discussed here is often referred to as
\textit{direct} RPA in quantum chemistry literature. In some literature, RPA in fact corresponds to the time-dependent Hartree-Fock theory, where
the exchange terms are also included. Nowadays, RPA with exchange included is usually referred to as RPAX or \textit{full} RPA}.
A simple way to include the exchange processes and eliminate the self-correlation error is to \textit{antisymmetrize} the two-electron Coulomb integrals
within the rCCD formulation of RPA [cf.~Eq.~\ref{eq:Ec_rpa_rccd}]. By doing so, a second-order screened exchange (SOSEX) 
\index{second-order screened exchange (SOSEX)} term 
\cite{Freeman:1977,Grueneis/etal:2009,Paier/etal:2010} is added, and the resultant RPA+SOSEX correlation energy is given by
\begin{equation}
	E_\text{c}^\text{RPA+SOSEX} = \frac{1}{2}\sum_{mn}^{occ}\sum_{ab}^{vir}\sum_{\sigma,\sigma'} 
 \left[B_{ma\sigma,nb\sigma'}-\delta_{\sigma\sigma'}B_{mb\sigma,na\sigma} \right] 
   T^\text{rCCD}_{nb\sigma',ma\sigma} \, ,
 \label{eq:Ec_RPA+SOSEX}
\end{equation}

with the two terms in Eq.~\ref{eq:Ec_RPA+SOSEX} corresponding to RPA and SOSEX \index{SOSEX} correlation energies respectively. 
The SOSEX contribution can be represented by the diagrams shown in the second row of Fig.~\ref{fig:rPT2_diagram}, where the leading term 
corresponds to the exchange contribution in PT2. The RPA+SOSEX scheme has the interesting feature that it is one-electron self-correlation free,
and improves substantially the total energy \cite{Grueneis/etal:2009,Paier/etal:2010}. The underbinding problem of RPA for chemically 
bonded molecules and solids is on average alleviated by the SOSEX correction.

As illustrated in Fig.~\ref{fig:rPT2_diagram}, both RPA and SOSEX correlations can be interpreted as 
\textit{infinite-order summations of selected types of diagrams}, with the PT2 terms in the leading order. 
This perspective is instructive for identifying important contributions that are still missing in the 
RPA+SOSEX scheme. In fact, at the second order, in addition to the direct and exchange terms, 
there is yet another type of contribution, arising from the SE, as already discussed above in the context of G{\"o}rling-Levy perturbation theory (Eqs.~\ref{eq:GL2} and \ref{eq:GL-SE}). For
beyond-RPA corrections, the SE-type correction was initially not derived according to the 
G{\"o}rling-Levy perturbation theory in Sec.~\ref{sec:GL2}, but rather the 
Rayleigh-Schr\"{o}dinger perturbation theory (RSPT) \cite{Szabo/Ostlund:1989}. 
As shown in Ref.~\cite{Ren/etal:2011},
within RSPT, the 2nd-order SE contribution is given by
  \begin{equation} 
	  \begin{split}
      E_\text{c}^\text{SE} = & \sum_{m}^{occ}\sum_{a}^{vir} \sum_{\sigma}
               \frac{|\langle\Phi_0|\hat{H}'|\Phi_{m\sigma}^{a\sigma}\rangle|^2}{E_0 - E^{(0)}_{m\sigma,a\sigma}} \nonumber \\
          =  & \sum_{m}^{occ}\sum_{a}^{vir}\sum_{\sigma}\frac{|\langle \psi_{m\sigma} | \hat{v}^\text{HF} - \hat{v}^\text{MF}|\psi_{a\sigma} \rangle |^2}
                  {\epsilon_{m\sigma} - \epsilon_{a\sigma}} \\
          =  & \sum_{m}^{occ}\sum_{a}^{vir}\sum_{\sigma}\frac{|\langle \psi_{m\sigma} | \hat{f} |\psi_{a\sigma} \rangle |^2}
                  {\epsilon_{m\sigma} - \epsilon_{a\sigma}} 
	 \end{split}
     \label{eq:Ec_SE}
  \end{equation}
where $\hat{v}^\text{MF}$ is the KS mean-field (MF) potential, according to which the reference state is generated, and $\hat{v}^\text{HF}$ is the non-local Hartree-Fock potential, evaluated 
using orbitals $\{\psi_{p\sigma}\}$ which themselves are generated according to the potential $\hat{v}^\text{MF}$. Furthermore,
$\hat{f}=-\nabla^2/2+\hat{v}^\text{ext} + \hat{v}^\text{HF}$ is the single-particle Hartree-Fock Hamiltonian (also known as the Fock matrix in the quantum chemistry literature), and
$\Phi_0$ and $\Phi_{m\sigma}^{a\sigma}$ are the ground-state Slater determinant formed 
by the lowest-occupied $\psi_{p\sigma}$'s  and the singly-excited configuration generated by
exciting one electron from an occupied state $m$ (with spin $\sigma$ and energy $\epsilon_{m\sigma}$) to a virtual state 
$a$ (with the same spin and energy $\epsilon_{a\sigma}$), respectively.
A detailed derivation of Eq.~\eqref{eq:Ec_SE}  can be found in the 
supplementary material of Ref.~\cite{Ren/etal:2011}.  
Obviously, for a Hartree-Fock reference where $\hat{v}^\text{MF}=\hat{v}^\text{HF}$, 
Eq.~\ref{eq:Ec_SE} becomes zero, a fact known as Brillouin theorem \cite{Szabo/Ostlund:1989}. 
Therefore, this term is not present in standard MP2 \index{MP2} theory based on the Hartree-Fock \index{Hartree-Fock} reference.  

One may notice that the SE expression derived within RSPT is different from the GL-SE given by Eq.~\ref{eq:GL-SE}, with the exact-exchange optimized effective potential (OEP) in the GL-SE is replaced in Eq.~\ref{eq:Ec_SE} by the mean-field potential, on which the reference state is based. In fact, RSPT can also be interpreted from the ACFDT perspective
along a linear adiabatic connection  path that linearly interpolates the non-interacting 
reference Hamiltonian and the full interacting Hamiltonian. However, in this case,
the electron density is not (and cannot be) fixed any longer along the adiabatic connection path.  Such a linear connection path was considered long ago by Harris and Jones in Ref.~\cite{Harris/Jones:1974}, 
and further discussed in Refs.~\cite{Toulouse/etal:2010,Aggelen/etal:2014,Klimes/etal:2015}.
As pointed by Klime\v{s} \textit{et al.} in Ref.~\cite{Klimes/etal:2015}, the SE term given by
Eq.~\ref{eq:Ec_SE} accounts for the change of the mean-field Hartree and exchange energy 
as $\lambda$ goes from 0 to 1, stemming from the change of both the electron density and
density matrix. In contrast, the GL-SE only accounts for the change of exchange energy, since the
electron density is fixed, and so does the Hartree energy.

From the practical point of view, Eq.~\ref{eq:Ec_SE} is easier to implement, since one can
use the readily available LDA- or GGA-type KS potential as $v^\mathrm{MF}$, without solving
a rather cumbersome OEP equation at the exact change level. Yet, as discussed above, more
physical effects missing in the standard RPA are accounted for by Eq.~\ref{eq:Ec_SE}. Due to
these advantages, the SE expression derived within RSPT is \textit{de facto} the SE
correction term for beyond-RPA schemes.

In Ref~\cite{Ren/etal:2011}, it was shown that adding the SE term of Eq.~\ref{eq:Ec_SE} to
RPA significantly improves the accuracy of vdW-bonded molecules, which the standard RPA
scheme generally underbinds, and the SOSEX \index{SOSEX} correction does not appreciably improve the situation. 
Similar to the RPA and SOSEX cases,
one can identify a sequence of single-excitation processes up to infinite order, as illustrated in the
third row in Fig.~(\ref{fig:rPT2_diagram}). Summing these single-excitation diagrams to infinite order
represents a renormalization of the 2nd-order SE, and is therefore termed as renormalized single excitations (rSE) \index{renormalized single excitations (rSE)} 
\cite{Ren/etal:2012b, Ren/etal:2013}.  Remarkably, this infinite summation can be translated into a closed expression similar to Eq.~(\ref{eq:Ec_SE}),
  \begin{equation}
       E_\text{c}^\text{rSE} = \sum_{m}^{occ}\sum_{a}^{vir} \sum_{\sigma}
       \frac{|\tilde{f}_{ma}^\sigma|^2}{\tilde{\epsilon}_{m\sigma} - \tilde{\epsilon}_{a\sigma}}\, ,
            \label{eq:E_rSE_final}
  \end{equation}
   where $\tilde{\epsilon}_{m\sigma}$ and $\tilde{\epsilon}_{a\sigma}$ are the eigenvalues obtained by diagonalizing separately the occupied-occupied and
   virtual-virtual subblocks of the Fock matrix, 
      \begin{equation}
       \begin{aligned}
        \sum_{l} f_{ml}^\sigma{\cal O}_{ln}^{\sigma} & = \tilde{\epsilon}_{m\sigma} {\cal O}_{mn}^\sigma \\
	\sum_{c} f_{ac}^\sigma{\cal U}_{cb}^{\sigma} & = \tilde{\epsilon}_{a\sigma} {\cal U}_{ab}^\sigma \, ,
       \end{aligned}
       \label{eq:semi_canonicalization}
       \end{equation}
  with $f_{pq}^\sigma = \langle \psi_{p\sigma} | \hat{f} | \psi_{q\sigma} \rangle$. Note that $f^\sigma_{pq}$ matrix is not diagonal since $\psi_{p\sigma}$'s are the reference KS orbitals
  and not the eigenfunctions of $\hat{f}$ -- the single-particle Hartree-Fock Hamiltonian. Now, $\tilde{f}_{ma}^{\sigma}$ in the 
  numerator of Eq.~(\ref{eq:E_rSE_final}) are the ``transformed" off-diagonal block of the Fock matrix
	       \begin{equation}
	           \tilde{f}_{ma}^{\sigma} = \sum_{n}^{occ}\sum_b^{vir}{\cal O^\ast}_{mn}^{\sigma} f_{nb}^{\sigma} {\cal U}_{ba}^{\sigma} \, ,
	    \end{equation}
 where the eigenvectors obtained in (\ref{eq:semi_canonicalization}) are used here as the transformation coefficients.
 The physical origin of the SE corrections is that the commonly used KS references for RPA and beyond-RPA calculations are not the optimal starting point.  
 The SE corrections accounts for the ``orbital relaxation" effect, which leads to a lowering of the ground-state total energy.
 Furthermore, it was found that SE and rSE corrections generally increases the binding energies, in particularly for weakly bonded systems.
 When the SE contributions are added, the electron density effectively contracts compared to that yielded by local and semilocal DFAs, and this
 reduces the Pauli repulsion between closed-shell atoms and molecules. Consequently, the underbinding behavior of the standard RPA is corrected
 for both weakly bonded molecules \cite{Ren/etal:2011,Ren/etal:2013} and solids \cite{Klimes/etal:2015}. Recently, the concept of rSE has been
 extended to Green's function by Yang and coworkers \cite{Jin/Su/Yang:2019,Li/Yang:2022}. 
 These authors showed that, by utilizing the so-called 
 renormalized singles Green's function, the starting point dependence in the usual perturbative $G_0W_0$-type calculations are significantly reduced \cite{Jin/Su/Yang:2019,Li/Yang:2022}.  

Diagrammatically, RPA, SOSEX and rSE \index{rSE} are three distinct infinite series of many-body terms, in which the three leading
terms correspond to the three terms in second-order many-body perturbation theory.  Summing them up,
the resultant RPA+SOSEX+rSE scheme can be viewed a \textit{renormalization} of 
the  second-order many-body perturbation theory. Therefore the RPA+SOSEX+rSE scheme is
also termed as  ``\textit{renormalized second-order perturbation theory}" or rPT2 \index{rPT2}.
Independent of the rPT2 \index{rPT2} scheme described above, Bates and Furche developed a beyond-RPA formalism termed as RPA 
renormalized many-body perturbation theory \cite{Bates/Furche:2013}. The essence of this formalism is to express the correlation energy
in terms of an integration over the polarization propagator (closely related to the density response function) along the AC path. The correlation energy can 
be improved by correcting the polarization propagator based on a series expansion in terms of the RPA polarization propagator multiplied with a four-point
kernel. Benchmark calculations \cite{Chen/etal:2018} show that the approximate exchange kernel (AXK) scheme within this formalism performs
better than RPA+SOSEX discussed above. However, the rSE \index{rSE} contribution is not included in the AXK scheme.

\section{Implementation of the fifth-rung functionals}
Now it is clear that the key in the calculations of fifth-rung functionals is to evaluate the exact-exchange energy (Eq.~\ref{eq:E_HF_x}) 
and the corresponding correlation energies that are formed explicitly with unoccupied KS orbitals, 
like the PT2 correlation used in DHAs (Eq.~\ref{eq:PT2}) and the RPA correlation in RPA-based methods (Eq.~\ref{eq:Ec_RPA}).  
The algorithms for evaluating the exact-exchange energy in the present context are exactly the same as 
that of the Hartree-Fock exchange energy, which are routinely done
in quantum chemistry calculations \cite{g09,TURBOMOLE}. The Hartree-Fock exchange is also the key component of hybrid density functionals, 
available in increasingly more softwares that can deal with periodic systems \cite{Paier/etal:2006,Levchenko/etal:2015,Lin/Ren/He:2020,Lin/Ren/He:2021}.
Here, we should not discuss the implementation of the Hartree-Fock exchange, but rather focus on the implementation of the correlation part of 
DHAs and RPA.

\subsection{Implementation of the DHAs}
Taking the XYG3-type DHAs using B3LYP orbitals (xDH@B3LYP, Eq.~\ref{eq:xdh@b3lyp}) as example, the PT2 correlation is apparently the most 
time comsuming part. The standard sum-over-state formula of the PT2 correlation was already in Eq.~\ref{eq:PT2}. For convenience, 
here we rewrite the PT2 correlation energy by separating out the \textit{direct} and \textit{exchange} terms,
\begin{equation}
	\label{eq:PT2_qc}
	E_{c}^\mathrm{PT2}=
	\frac{1}{2}\sum_{mn}^{occ}\sum_{ab}^{vir}\sum_{\sigma,\sigma'}\left(ma,\sigma|nb,\sigma'\right)
	\left[\frac{\left(am,\sigma|bn,\sigma'\right)-\left(bm,\sigma|an,\sigma'\right)\delta_{\sigma\sigma'}}
	{\varepsilon_{m\sigma}+\varepsilon_{n\sigma'}-\varepsilon_{a\sigma}-\varepsilon_{b\sigma'}}\right],
\end{equation}
with $\left(ma,\sigma|nb,\sigma'\right)=\langle \psi_{m\sigma}\psi_{n\sigma'}|\psi_{a\sigma}\psi_{b\sigma'}\rangle$ being two-electron Coulomb repulsion integrals for molecular orbitals $\left\{ \psi_{p\sigma} \right\}$.
This repulsion integrals can be further expressed in terms of basis functions $\{\phi_i(\bfr)\}$
\begin{equation}
	\label{eq:ao2mo_eri}
	\left(ma,\sigma|nb,\sigma'\right)=\sum_{ijkl}\left( \phi_i\phi_j|\phi_k\phi_l \right)c_{m\sigma}^{i*}c_{a\sigma}^{j}c_{n\sigma'}^{k*}c_{b\sigma'}^{l},
\end{equation}
where $\left( \phi_i\phi_j|\phi_k\phi_l \right)$ are the two-electron integrals for the atomic orbitals, 
and $c_{m\sigma}^{i}$ are the eign-coefficients for the molecular orbitals.

Here we briefly describe the PT2 implementation in the FHI-aims code \cite{Blum/etal:2009,Ren/etal:2012}, which employs real atom-centered numeric orbitals 
(NAOs) as basis functions,
\begin{equation}
   \phi_i(\bfr) = u_k(r)Y_{lm}(\theta,\phi)
\end{equation}
with $u_k(r)$ being the numerically tabulated radial function and $Y_{lm}(\theta,\phi)$ the spherical harmonics. Unlike the gaussian-type 
orbital (GTO) basis functions, NAOs does not have analytical algorithms to evaluate $\left( \phi_i\phi_j|\phi_k\phi_l \right)$ integrals efficiently.
Therefore, the so-called RI technique was employed, which represents the pair products of real atomic basis functions 
$\phi_i^*(\bfr)\phi_j(\bfr)=\phi_i(\bfr)\phi_j(\bfr)$ in terms of auxiliary basis functions, 
\begin{equation}
	\phi_i(\bfr)\phi_j(\bfr) \approx \sum_{\mu}C_{ij}^{\mu}P_{\mu}(\bfr).
	\label{eq:RI_expansion}
\end{equation}
Our auxiliary basis functions are also constructed as a numerically tabulated radial function multiplied by spherical harmonics,
          \begin{equation}
			  \label{eq:aux_bas}
             P_\mu(\bfr) = {\xi}_s(r)Y_{lm}(\theta,\phi)
          \end{equation}
but the radial function ${\xi}_s(r)$ has different shapes from $u_k(r)$. In fact, they are generated in order to best represent the pair products of the
one-electron orbitals $\{\phi_i\phi_j \}$. Details on how the auxiliary basis functions are constructed can be found in Refs.~\cite{Ren/etal:2012,Ihrig/etal:2015}.
         
Once the auxiliary basis functions $\{P_\mu(\bfr)\}$ are constructed, one can start to determine the triple expansion coefficients $C_{ij}^\mu$. For 
any finite auxiliary basis set, the expansion in Eq.~\ref{eq:RI_expansion} is an approximation, incurring an error 
$\delta \rho_{ij}(\bfr) = \sum_{\mu} C_{ij}^\mu P_\mu(\bfr) - \phi_i^\ast(\bfr) \phi_j(\bfr)$. The accuracy of this approximation will not only depend
on the quality and size of the auxiliary basis, but also depend on the expansion coefficients $C_{ij}^\mu$. In the RI approach with 
Coulomb metric \cite{Vahtras/Almlof/Feyereisen:1993}, instead of minimizing the norm of the error $\left(\delta \rho_{ij} | \delta \rho_{ij} \right)$, 
one minimizes the self Coulomb repulsion of 
this error $\left(\delta \rho_{ij} |v| \delta \rho_{ij} \right)$, leading to the following expression for $C_{ij}^\mu$,
          \begin{equation}
              C_{ij}^\mu = \sum_{\nu} (ij|v|\nu) V^{-1}_{\nu\mu}
          \end{equation}
where 
          \begin{equation}
             (ij|v|\nu) = \iint d\bfr d\bfrp \frac{\phi_i^*(\bfr)\phi_j(\bfr) P_\nu(\bfrp)}{|\bfr-\bfrp|}\, ,
          \end{equation}
	  and $V^{-1}$ is the inverted Coulomb matrix with elements given by
	       \begin{equation}
		       V_{\mu\nu}=\iint d\bfr d\bfrp \frac{P_\mu(\bfr) P_\nu(\bfrp)}{|\bfr-\bfrp|}
                      \label{eq:V_matrix}
	       \end{equation}
where $N_\mathrm{aux}$ is the number of auxiliary basis functions. 
With such kind of RI expansion, the two-electron integrals can be approximated as
\begin{equation}
	\label{eq:ri_v}
	\left( \phi_i\phi_j|\phi_k\phi_l \right) \approx \sum_{\mu\nu}C_{ij}^{\mu}V_{\mu\nu}C_{kl}^{\nu} = \sum_{\mu}M_{ij}^{\mu}M_{kl}^{\mu},
\end{equation}
where a new rank-3 tensor $M_{ij}^{\mu}$ is defined as,
\begin{equation}
	M_{ij}^{\mu}=\sum_{\nu}(ij|\nu)V_{\mu\nu}^{-1/2}=\sum_{\nu}C_{ij}^{\nu}V_{\nu\mu}^{1/2}\, .
\end{equation}
By using the RI expansion (Eq.~\ref{eq:ri_v}), the two-electron Coulomb repulsion integrals (Eq.~\ref{eq:ao2mo_eri}) can be decoupled as
follows
\begin{equation}
	\left(ma,\sigma|nb,\sigma'\right)\approx \sum_{\mu,\nu}O_{ma,\sigma}^{\mu}V_{\mu\nu}O_{nb,\sigma'}^{\nu} = \sum_{\mu}Q_{ma,\sigma}^{\mu}Q_{nb,\sigma'}^{\mu},
\end{equation}
with the three-orbital integrals $O_{ma,\sigma}^{\mu}$ defined as
\begin{equation}
	O_{ma,\sigma}^{\mu}=\sum_{ij}C_{ij}^{\mu}c_{m\sigma}^{i*}c_{a\sigma}^{j}, ,
        \label{eq:MO_expan_coefficients}
\end{equation}
and $Q_{ma,\sigma}^{\mu}$ defined as
\begin{equation}
	Q_{ma,\sigma}^{\mu}=\sum_{ij}M_{ij}^{\mu}c_{m\sigma}^{i*}c_{a\sigma}^{j}\, .
       \label{eq:MO_expan_coefficients1}
\end{equation}
As a result, the RI version of the PT2 correlation energy (RI-PT2) is given by
\begin{equation}
	\label{eq:PT2_RI}
	\begin{split}
	    E_{c}^\mathrm{PT2}=&
	    \frac{1}{2}\sum_{mn}^{occ}\sum_{ab}^{vir}\sum_{\sigma,\sigma'}\left( \sum_{\mu}O_{ma,\sigma}^{\mu}Q_{nb,\sigma'}^{\mu} \right)
	    \left[\frac{\sum\limits_{\mu}Q_{ma,\sigma}^{\mu}Q_{nb,\sigma'}^{\mu}-
		\sum\limits_{\mu}Q_{bm,\sigma}^{\mu}Q_{an,\sigma'}^{\mu}\delta_{\sigma\sigma'}}
	{\varepsilon_{m\sigma}+\varepsilon_{n\sigma'}-\varepsilon_{a\sigma}-\varepsilon_{b\sigma}}\right]\, .
	\end{split}
\end{equation}
The RI-PT2 implementation (Eq.~\ref{eq:PT2_RI}) scales as $O(N^5)$ with respect to the size of the system. Although it has the same scaling
exponent as the standard PT2 formula (Eq.~\ref{eq:PT2_qc}), the prefactor in RI-PT2 is one to two orders of magnitude smaller than in the
standard PT2~\cite{Ren/etal:2012,Ihrig/etal:2015}. In recent years, lower-scaling algorithms of PT2 have also been developed, which greatly
enhanced the applicability of PT2-based methods, in particular DHAs, to large systems \cite{takatsuka:2008,lee:2020,zhang:2021D}.

This conventional (global) RI approach works extremely well for small molecules. For big molecules and periodic systems, one may resort to 
a localized variant of the RI approach \cite{Ihrig/etal:2015}. With enhanced auxiliary basis sets, the localized RI approach has been
shown to be sufficiently accurate in practical calculations \cite{Ihrig/etal:2015,Lin/Ren/He:2020,Ren/etal:2021}, and is instrumental 
for periodic systems \cite{Levchenko/etal:2015}. The RI version of the PT2 correlation energy (Eq.~\ref{eq:PT2_RI}) can be easily generalized 
from finite molecules to periodic systems. Taking advantage of the localized RI techniques, one can achieve excellent massive-parallel efficiency
in periodic PT2 implementation as well. It allows one to perform a comprehensive benchmark on the 
numerical convergence of the PT2 correlation 
energies with respect to the basis set and $\bfk$-mesh sizes \cite{IgorZhang/etal:2019}, which is crucial to make the periodic implementation of
DHAs practical \cite{zhang:2021C}.

\subsection{Implementation of the RPA method}
As discussed in Sec.~\ref{sec:ACFDT_RPA}, in most practical RPA calculations, the evaluation of the RPA XC energy $E_\mathrm{xc}^\mathrm{RPA}$ as given in Eqs.~\ref{eq:E_xc_RPA}-\ref{eq:Ec_RPA}
is done perturbatively employing the KS orbitals and orbital energies generated by a preceding KS-DFT calculation with certain lower-rung functionals. 
The PBE-GGA is often used to serve this purpose, and the RPA calculation based a PBE reference is commonly
denoted as ``RPA@PBE''. As already indicated in Eq.~\ref{eq:RPA_E_total}, in standard RPA@PBE calculations, the ground-state total energy is given by
   \begin{equation}
	   \begin{split}
	   E^\mathrm{RPA@PBE} &= E^\mathrm{PBE} - E_\mathrm{xc}^\mathrm{PBE} + E_\mathrm{xc}^\mathrm{RPA@PBE} \nonumber \\
			      &= T_s^\mathrm{PBE} + E_\mathrm{ext}^\mathrm{PBE} + E_\mathrm{H}^\mathrm{PBE} +  E_\mathrm{x}^\mathrm{EX@PBE} +
			         E_\mathrm{c}^\mathrm{RPA@PBE} \nonumber \\
			     & = \langle \Phi_\mathrm{PBE} |\hat{H}|\Phi_\mathrm{PBE}\rangle + E_\mathrm{c}^\mathrm{RPA@PBE} \,
			 \end{split}
           \label{eq:E_tot_RPA-PBE}
   \end{equation}
where $T_s^\mathrm{PBE}, E_\mathrm{ext}^\mathrm{PBE}, E_\mathrm{H}^\mathrm{PBE}$, $E_\mathrm{xc}^\mathrm{PBE}$ are the 
noninteracting kinetic energy, the external potential energy, the Hartree energy, and the XC energy obtained from the self-consistent PBE calculation, respectively.
The RPA@PBE total energy defined in Eq.~\ref{eq:E_tot_RPA-PBE} is obtained by subtracting the PBE XC energy from the PBE total energy, and then adding 
the RPA XC energy, evaluated using PBE orbitals and orbital energies. Eq.~\ref{eq:E_tot_RPA-PBE} also suggests that the RPA total energy can be seen 
as the sum of the Hartree-Fock energy evaluated with respect to the PBE reference and the RPA@PBE correlation energy. 

To develop efficient algorithms to evaluate Eq.~\ref{eq:Ec_RPA}, the key is to realize that both $\chi^0$ and $v$ are non-local operators,
depending on two coordinates in space.
Their real-space forms $\chi^0(\bfr,\bfrp,i\omega)$ and $v(\bfr,\bfrp)$ can be seen as basis representations of the corresponding operator $\chi^0$ and $v$
in terms of real-space grid points. Under such a discretization, $\chi^0$ and $v$ become matrices of dimension as large as the number of 
real space grid points. This is not an efficient representation since the number of grid points is rather large, especially in all-electron implementations.
To deal with this problem, an auxiliary basis set $\{P_\mu(\bfr) \}$, in the form given by Eq.~\ref{eq:aux_bas},
is introduced to represent 
$\chi^0$ and $v$, namely,
\begin{equation}
    \chi^0(\bfr,\bfrp,i\omega) = \sum_{\mu,\nu}^{N_\mathrm{aux}} P_\mu(\bfr) \chi^0_{\mu\nu}(i\omega) P_\nu(\bfrp),
    \label{eq:chi0_expan}
\end{equation}
and 
the $V_{\mu\nu}$ matrix which has been defined in Eq.~\ref{eq:V_matrix}. 
To find the matrix form $\chi^0_{\mu,\nu}(i\omega)$, one needs to expand the products of KS orbitals in terms of $P_\mu(\bfr)$;
from Eqs.~\ref{eq:RI_expansion} and \ref{eq:MO_expan_coefficients}, one has 
	       \begin{equation}
	            \psi_{p\sigma}^\ast(\bfr) \psi_{q\sigma}(\bfr) = \sum_{\mu} O_{pq,\sigma}^\mu P_\mu(\bfr)\, .
                \label{eq:RI_expansion_MO}      
	       \end{equation}
Inserting Eq.~\ref{eq:RI_expansion_MO} into Eq.~\ref{eq:indep_response}, and
comparing the resultant expression to Eq.~\ref{eq:chi0_expan}, one arrives at
              \begin{equation}
		      \chi^0_{\mu\nu}(i\omega)=\sum_{p,q,\sigma} \frac{(f_{p\sigma} - f_{q\sigma}) O_{pq,\sigma}^\mu O_{pq,\sigma}^{\nu\ast}}{\epsilon_{p\sigma} - \epsilon_{q\sigma} - i\omega}\, .
		      \label{eq:chi0_mat_ABF}
	      \end{equation}
Thus, within the auxiliary basis representation, $\chi^0$ and $v$ become $N_\mathrm{aux} \times N_\mathrm{aux}$ matrices.
Since $N_\mathrm{aux}$ is typically much smaller than the number of real-space grid points, or the number of pair products
of KS orbitals, the $\chi^0$ matrix in Eq.~\ref{eq:chi0_mat_ABF} and the V matrix in Eq.~\ref{eq:V_matrix}
    can be seen as compressed representation of these operators, with the trace of their product unchanged,
	       \begin{equation}
		       \textrm{Tr}\left[\chi^0 (i\omega) v\right] = \iint \chi^0(\bfr,\bfrp,i\omega)v(\bfrp,\bfr) d\bfr d\bfrp
			 = \sum_{\mu\nu} \chi^0_{\mu\nu}(i\omega)V_{\mu\nu}\, .
			 \label{eq:trace_chi0v}
	       \end{equation}
Based on this observation, one can conclude that the computed RPA correlation energy is independent of the basis representation of $\chi^0$ and $v$
operators. Thus one may equivalently interpret Eq.~\ref{eq:Ec_RPA} as matrix algebra with $\chi^0$ and $v$ represented within the auxiliary basis 
as given by Eqs.~\ref{eq:chi0_mat_ABF} and \ref{eq:V_matrix}. 
Using the property $\mathrm{Tr}(\mathrm{ln}(A)=\mathrm{ln}(\mathrm{Det}(A))$ ($\mathrm{Det}(A)$ being the determinant of the matrix $A$), 
one obtains the final working expression for evaluating the RPA correlation energy,
              \begin{equation}
				  \begin{split}
		      E_\text{c}^\text{RPA} & = \frac{1}{2\pi}\int_0^\infty d\omega \mathrm{ln} \left[ \mathrm{Det}(1-\chi^0 (i\omega) v) + 
						          \chi^0 (i\omega) v \right] \\
		                          & = \frac{1}{2\pi}\int_0^\infty d\omega \mathrm{ln} \left[ \mathrm{Det}(1-\Pi (i\omega)) + 
						          \Pi(i\omega)  \right] \, .
				 \end{split}
  		      \label{eq:Ec_RPA_final}
	      \end{equation}
In Eq.~\ref{eq:Ec_RPA_final}, we have introduced an intermediate quantity $\Pi(i\omega) = V^{1/2} \chi^0(i\omega) V^{1/2}$ and used the property
$\mathrm{Tr}[{V^{1/2}\chi^0V^{1/2}]} = \mathrm{Tr}[\chi^0 V]$. It is easy to recognize that the $\Pi$ matrix can be directly evaluated as
              \begin{equation}
		      \Pi^0_{\mu\nu}(i\omega)=\sum_{p,q,\sigma} \frac{(f_{p\sigma} - f_{q\sigma}) Q_{pq,\sigma}^\mu Q_{pq,\sigma}^{\nu\ast}}{\epsilon_{p\sigma} - \epsilon_{q\sigma} - i\omega}\, .
		      \label{eq:Pi_mat_ABF}
	      \end{equation}
where the coefficients  $Q_{pq,\sigma}^\mu$ are defined in Eq.~\ref{eq:MO_expan_coefficients1}.

The frequency integration in Eq.~\ref{eq:Ec_RPA_final} can be done relatively
easily, since the integrand is rather smooth and peaked at low frequency values. A modified Gauss-Legendre grid, which transforms a standard 
Gauss-Legendre grid in the range $[-1,1]$ to $[0,\infty]$,
              \begin{equation}
                     \tilde{x}_i  =  x_0 \frac{1+x_i}{1-x_i}, ~~~~~ \tilde{w}_i  =  w_i \frac{2x_0}{(1-x_i)^2} 
                   \label{eq:mod_GL_grid}
              \end{equation}
works rather well for most systems. In Eq.~\ref{eq:mod_GL_grid}, $(x_i, w_i)$ are the abscissas and weights of the grid points generated from 
the Gauss-Legendre quadrature formula in an integration range $[-1,1]$, whereas $(\tilde{x}_i, \tilde{w}_i)$ are abscissas and weights of
the transformed grid.  Usually a few tens of frequency grid points are sufficient to get highly accurate results, except for systems 
with vanishing gaps, where considerably more grid points have to be used. Alternative frequency grids have been developed for evaluating
the RPA correlation energies \cite{Eshuis/Yarkony/Furche:2010,Kaltak/Klimes/Kresse:2014,Kaltak/etal:2014b}, 
which have been shown to work well for small gap systems. Especially the minimax grid has proved to be very efficient when
used for the Fourier transform between the frequency and time domains, which is essential for the success of the cubic-scaling space-time algorithm 
for the RPA correlation energy calculations \cite{Kaltak/Klimes/Kresse:2014,Kaltak/etal:2014b}.

The implementation scheme described above is known as the RI approach to RPA \cite{Eshuis/Yarkony/Furche:2010,Ren/etal:2012}. 
From the above discussion, one may see that once the matrix forms of $\chi^0$ and $v$ within the auxiliary basis are determined, 
the rest of the RPA calculation is rather straightforward. The key steps in RPA calculations are thus: 1) Generate the auxiliary 
basis functions $\{P_\mu(\bfr)\}$; 2) Determine the RI coefficients $O_{pq,\sigma}^\mu$; and 3) Construct the $\chi^0_{\mu,\nu}$ 
matrix using (\ref{eq:chi0_mat_ABF}), or alternatively the $\Pi$ matrix using Eq.\ref{eq:Pi_mat_ABF}. The choice of auxiliary basis functions $\{P_\mu(\bfr)\}$
depends on the underlying one-electron basis functions used in the KS-DFT calculations. 
The above-outlined algorithm works rather well for molecular geometries. This algorithm has been extended to periodic systems, utilizing the
localized RI approximation. Periodic RPA implementation based on localized RI within the NAO basis set framework has also been available in 
FHI-aims \cite{Blum/etal:2009,Ren/etal:2012} and used in
production calculations \cite{IgorZhang/etal:2019,Yang/Ren:2022}. The implementation details, which has not been published yet, follow
similar algorithms as for the periodic $GW$ case, which has been described in Ref.~\cite{Ren/etal:2021}.

The key step in RPA correlation energy calculations is to construct the $\chi^0$ matrix in Eq.~\ref{eq:chi0_mat_ABF}. 
Since the number of auxiliary basis functions $N_\mathrm{aux}$ scales linearly with
the number of one-electron basis functions, the computational cost of Eq.~\ref{eq:chi0_mat_ABF} scales as $O(N^4)$ with respect
to the size of the system, and represent the computational bottleneck of RPA calculations. In recent years, lower-scaling algorithms 
have also been developed \cite{Kaltak/Klimes/Kresse:2014,Kaltak/etal:2014b,Beuerle/Ochsenfeld:2018}, which holds promise for extending
the applicability of RPA to large systems.

\section{Performance of the fifth-rung functionals and prototypical applications}
The increasing interest in DHAs and RPA in recent years comes not only from their appealing concept, but also from their remarkable performance 
in practical applications. 
First of all, DHAs and RPA can describe vdW \index{vdW} interactions in a seamless way \cite{Dobson:1994}, their performance has been 
demonstrated for varying kinds of weak-interacting systems \cite{Harl/Kresse:2008,Ren/etal:2011, zhang:2012B, zhang:2022A}.
For lower-rung DFAs ranging from LDA to hybrid functionals, the long-range vdW \index{vdW} 
interactions are not captured, and \textit{ad hoc} corrections have to be added in order to describe systems where vdW \index{vdW} interactions
play an important role. 
Furthermore, it seems that DHAs and RPA are able to describe the delicate energy differences in complex bonding situations, including molecules 
adsorbed on surfaces \cite{Ren/etal:2009,Schimka/etal:2010,zhang:2022A}, the isostructural phase transition \cite{Casadei/etal:2012}
and the relative stability of different polymorphs of crystals \cite{Lebegue/etal:2010,Sengupta/Bates/Ruzsinszky:2018,Zhang/Cui/Jiang:2018,
Cazorla/Gould:2019, Yang/Ren:2022,zhang:2021C}, the binding and ex-foliation energies of layered compounds \cite{Bjoerkman/etal:2012}, 
and the formation energy of defects \cite{Bruneval:2012,Kaltak/Klimes/Kresse:2014}. Finally, DHAs and RPA yield accurate chemical reaction barrier
heights \cite{Ren/etal:2012b,zhang:2009A,Zhang/Xu:2014}, which is crucial for reliably estimating the rate of chemical reactions. 

	\subsection{vdW interactions}
	vdW interactions arises from the coupling between spontaneous quantum charge fluctuations separated in space. For two well-separated, 
	spherically symmetric and chargely neutral systems, the interaction energy goes like
	   \begin{equation}
		   \Delta E \sim \frac{C_6}{R^6} ~~~~ \mathrm{for}~~ R\rightarrow \infty
              \label{eq:C6overR6}
	   \end{equation}
	   where $C_6$ is the dispersion coefficient and $R$ is the separation between the two subsystems. Dobson showed that the interaction energy
	   between two subsystems $A$ and $B$ obtained by RPA exactly follows the asymptotic behavior given by Eq.~\ref{eq:C6overR6}, 
	   with the RPA C$_6$ coefficients given by
	   \begin{equation}
		   C_6^\mathrm{RPA} = \frac{3}{\pi} \int d\omega \alpha^\mathrm{RPA}_A(i\omega) \alpha^\mathrm{RPA}_B(i\omega)\, 
              \label{eq:RPA_C6}
	   \end{equation}
	   where $\alpha^\mathrm{RPA}_A(i\omega)$ is the RPA polarizability of the subsystem $A$, which can be obtained by integrating over the
	   microscopic RPA response function as,
	   \begin{equation}
		   \alpha^\mathrm{RPA}_A(i\omega) = \frac{1}{3} \sum_{i=1}^3 \iint d\bfr d\bfrp r_i \chi^\mathrm{RPA}(\bfr,\bfrp,i\omega) r'_i
            \label{eq:RPA_polarisability}
	   \end{equation}
	   with $r_{1,2,3} = x,y,z$. In Ref.~\cite{Gao/Zhu/Ren:2020}, the RPA $C_6$ coefficients for a set of atoms and molecules 
were evaluated according to Eqs.~\ref{eq:RPA_C6} and \ref{eq:RPA_polarisability}. It was shown that, on average, the 
$C_6$ coefficients were underestimated by 13\% by RPA@PBE compared to reference values \cite{Gao/Zhu/Ren:2020}.

\begin{figure}[t]
 \centering
 \vskip 5mm
 \includegraphics[width=0.5\textwidth]{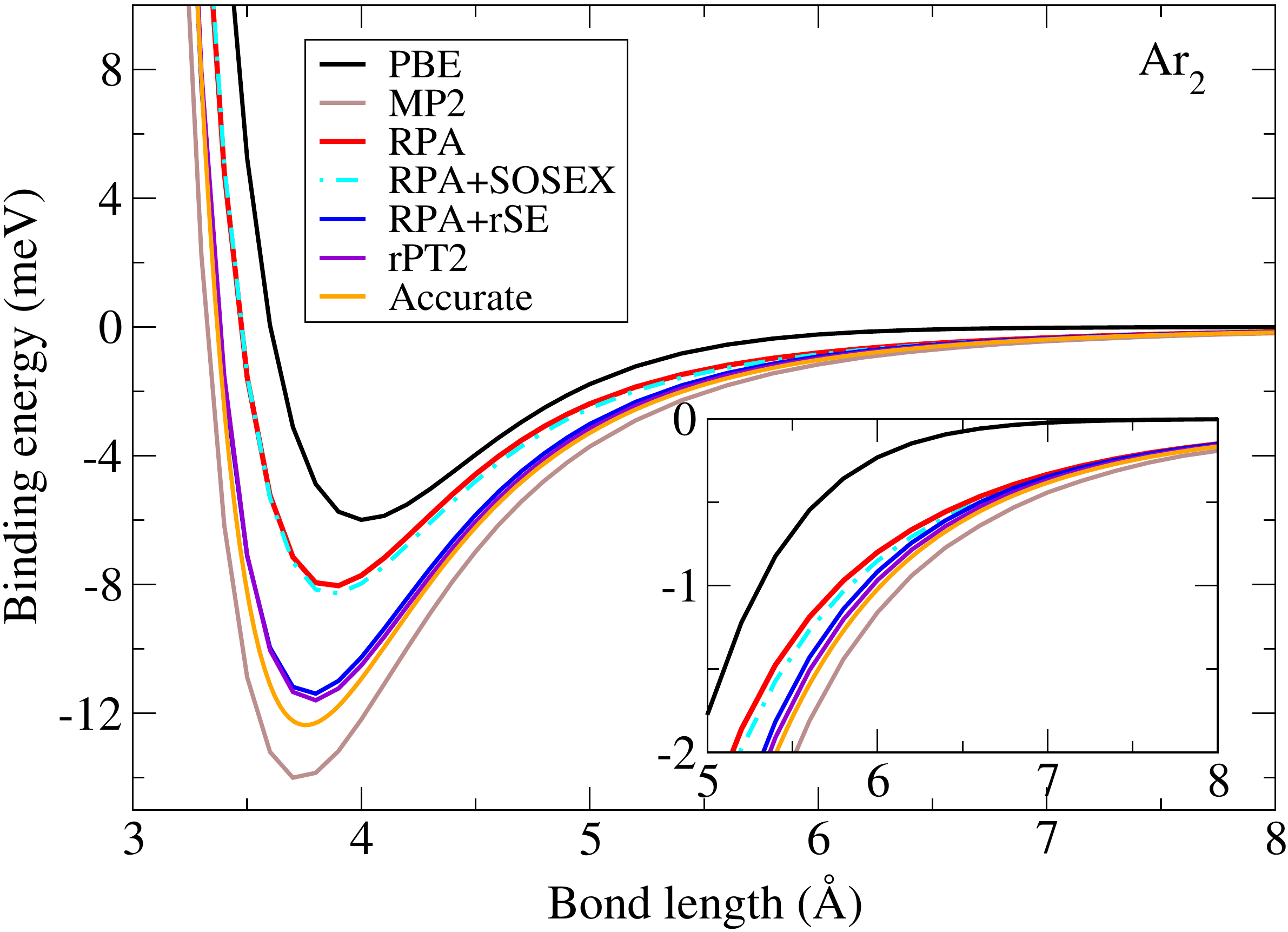}
 \caption{Binding energy curves for Ar$_2$ obtained using PBE, MP2, and
	    RPA-based methods.  All RPA-based methods use PBE orbitals as input. Calculations are done using the FHI-aims code \cite{Blum/etal:2009,Ren/etal:2012}. 
	    ``Accurate" reference curve is given by the Tang-Toennies potential model for Ar$_2$ \cite{Tang/Toennies:2003}. The
	     inset is a zoom-in of the asymptotic behavior at large separations.}
 \label{fig:Ar2_dissociation}
\end{figure}

The accuracy of the RPA $C_6$ coefficient only tells about the performance of the method in the asymptotic regime, but not in the entire bonding region. In Fig.~\ref{fig:Ar2_dissociation}, the full
binding energy curves of Ar$_2$ obtained using PBE and RPA-based methods are presented. In addition, the result of MP2, the simplest
post-Hartree-Fock quantum chemistry approach capable of describing vdW \index{vdW} interactions, is also included here for comparison. The reference curve
is given by the Tang-Toennies potential model, with model parameters determined from experiment. The PBE underestimates the bonding strength of Ar$_2$ considerably,
and this is appreciably improved by RPA. More importantly, at large separations, the PBE binding energy decays rapidly (in fact exponentially) to the energy zero,
but the RPA curve displays the correct $C_6/R^6$ asymptotic behavior. Compared to the reference result, however, the RPA still shows a substantial underbinding, in contrast to the MP2 result which shows too strong binding of Ar$_2$. The underbinding issue of RPA for vdW dimers arises from the too strong Pauli
repulsion, which is in turn due to the Hartree-Fock part of the RPA total energy, obtained with the semi-local GGA orbitals. As discussed in Sec.~\ref{sec:beyond_rpa}, this issue can be fixed by
the SE correction \cite{Ren/etal:2011}, and in particular its renormalized version \cite{Ren/etal:2013}.
As shown in Fig.~(\ref{fig:Ar2_dissociation}), the RPA+rSE \index{RPA+rSE} scheme brings the binding energy curve of Ar$_2$ in close agreement with the reference one. 
The SOSEX \index{SOSEX} correction, however, does not have a noticeable effect here. Thus the final rPT2 \index{rPT2} binding energy curve is almost on top of 
the RPA+rSE \index{RPA+rSE} one. 
The performance of RPA-based methods for other rare-gas dimers can be found in Ref.~\cite{Ren/etal:2013}.

\begin{figure}[t]
 \centering
	\begin{picture}(400,320)(0,0)
        \put(30,300){(a)}
	\put(0,220){\includegraphics[width=0.2\textwidth]{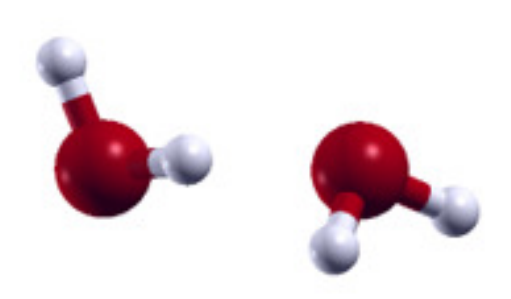}}
	\put(120,210){ \includegraphics[width=0.2\textwidth]{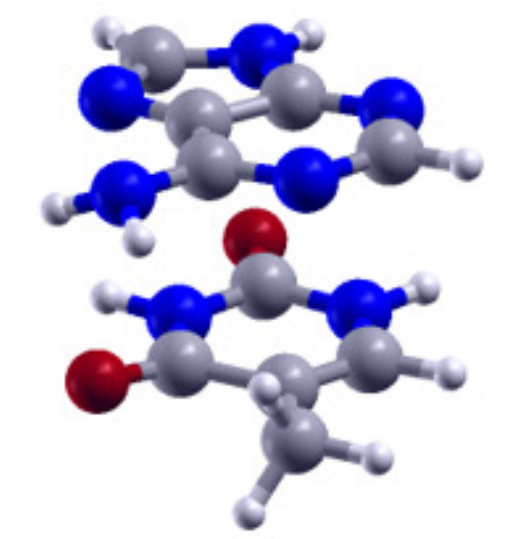}}
	\put(260,210){ \includegraphics[width=0.2\textwidth]{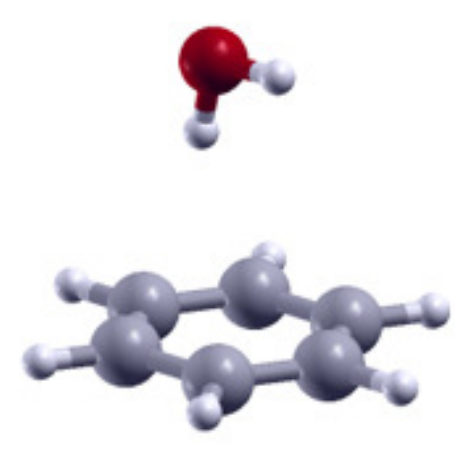}}
        \put(30,150){(b)}
	\put(60,0){ \includegraphics[width=0.5\textwidth]{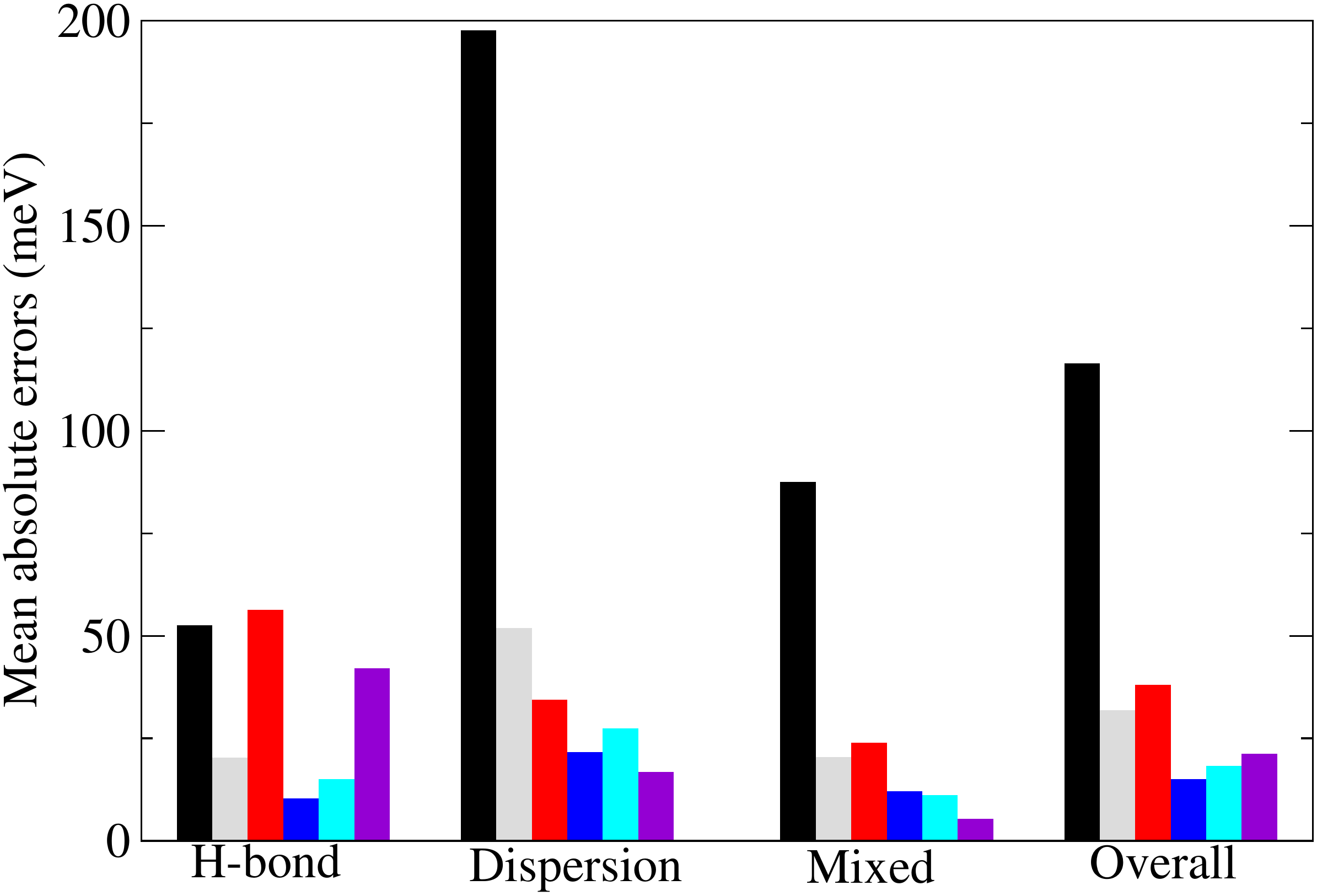}}
	\put(10,190){Water dimer}
	\put(120,190){Adenine-Thymine dimer}
	\put(260,190){Water-Benzene dimer}
	\end{picture}
	\caption{(a) Structures of water dimer, adenine-thymine dimer, and water-benzene dimer. C, O, N, and H atoms are represented by grey, red, blue, and white 
	 balls. (b) MAEs (in meV) for the S22 test set given by PBE, MP2, RPA, RPA+rSE, RPA+SOSEX, and rPT2 methods. The CCSD(T) results of 
	 Takatani {\it et al.} \cite{Takatani/etal:2010} are used here as the reference.}
 \label{fig:S22}
\end{figure}

A widely used benchmark set for weak interactions are the S22 test set designed by 
Jure\v{c}ka et al. \cite{Jurecka/etal:2006}. This test set collects 22 molecular dimers, among which
7 dimers are of hydrogen binding type, 8 of pure vdW (also called ``dispersion") bonding, and another 7 of mixed type. Figure~\ref{fig:S22}(a) 
shows the structures of water dimer, adenine-thymine dimer, and water-benzene dimer, representing the three bonding types, respectively.
Because of its good  representability and the availability of accurate reference interaction energies obtained
using the the CCSD(T) method \cite{Takatani/etal:2010}, S22 has been widely used for benchmarking the performance of or training the parameters for 
computational schemes that aim at describing weak interactions. The performance of RPA and some of the RPA-related methods has been benchmarked 
for this test set \cite{Zhu/etal:2010,Eshuis/Furche:2011,Ren/etal:2011,Eshuis/Furche:2012}.

In Fig.~\ref{fig:S22}(b) the mean absolute errors (MAE) of PBE, MP2, and RPA-based methods are presented for the three subsets separately and
for the entire S22 set. Both PBE and the correlated methods can describe the hydrogen bonding well, since this type of bonding is dominated by the
electrostatic interactions, which has already been captured by the semi-local functionals to a large extent. The error of the standard RPA method is still 
appreciable (MAE > 1 kcal/mol $\approx$ 43 meV) arising from its general underbinding behavior. This can again be corrected by rSE \index{rSE}, 
or by SOSEX \index{SOSEX} terms.
However, adding the two types of corrections together, the rPT2 \index{rPT2} scheme tends to overbind the hydrogen-bonded molecules, and hence the MAE increases again.
For dispersion and mixed bondings, RPA performs better, and the MAE can be further decreased by rSE \index{rSE} and SOSEX \index{SOSEX} corrections. The MP2 method, on the other hand, 
yields a relatively large MAE for dispersion-bonded molecules, owing to its well-known overbinding problem for this type of interaction. This benchmark
test indicates that the RPA+rSE scheme is a suitable approach which can be recommended 
for describing weak interactions. The advantage of this scheme is that it
does not noticeably increase the computational cost, compared to the standard RPA scheme.

\begin{figure}[t]
 \centering
 \vskip 5mm
 \includegraphics[width=0.6\textwidth]{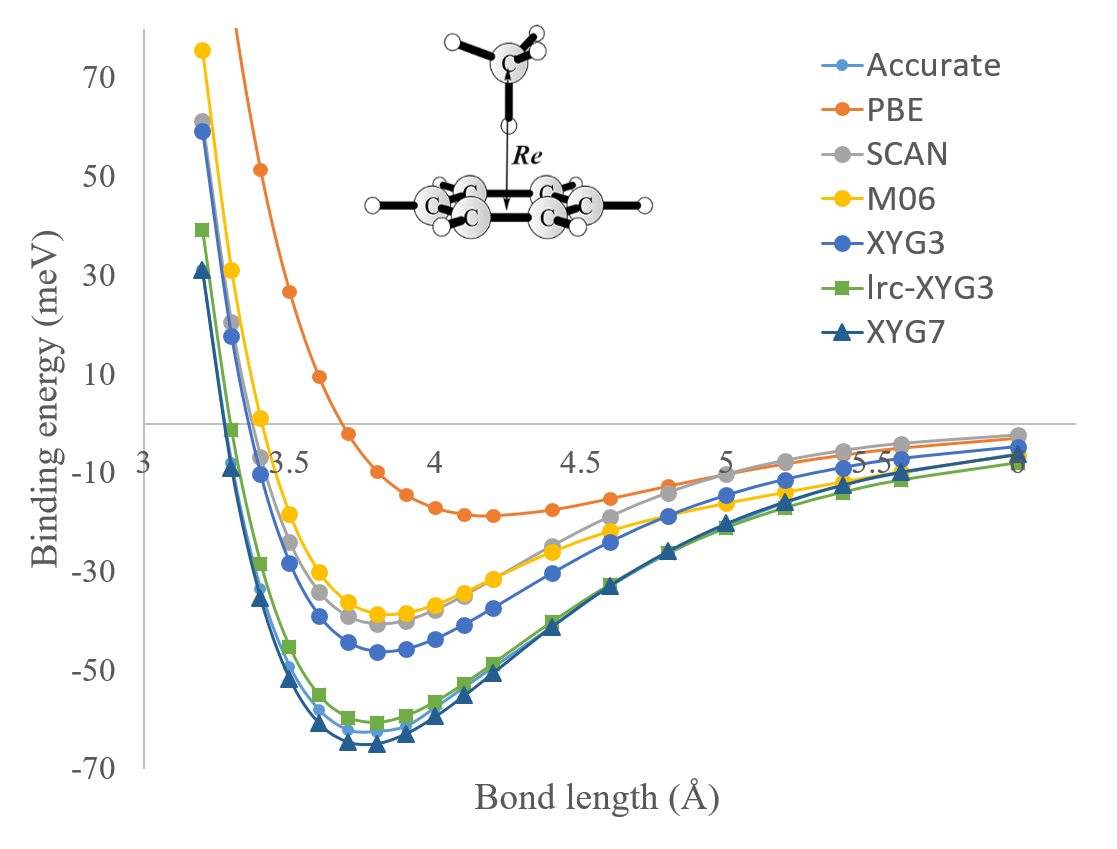}
 \caption{Binding energy curves for CH$_4$-C$_6$H$_6$ obtained using different kinds of popular methods in quantum chemistry or
 computational materials science. Results are collected from Ref.~\cite{zhang:2022A}.}
 \label{fig:MB_dissociation}
\end{figure}

PT2-based DHAs include a portion of advanced PT2 correlation, and thus also exhibit a notable improvement over conventional DFAs for vdW 
interactions. Taking the binding curve of methane and benzene as example (Fig.~\ref{fig:MB_dissociation}),  we see again that the PBE underestimates
the vdW interaction notably, yielding a too shallow potential well and too long bond length. 
The XYG3 performs much better than the PBE, and is even slightly better than the best meta-GGA SCAN functional \cite{Sun/Ruzsinszky/Perdew:2015} 
and the popular hybrid meta-GGA M06-2X functional \cite{zhao_new_2006}
in quantum chemistry. However, as shown in Fig.~\ref{fig:MB_dissociation}, the XYG3 still exhibits an underbinding tendency.
In fact, extensive benchmarks suggest that empirical dispersion correction is still needed for most of existing DHAs
\cite{goerigk_look_2017}. By using the long-range part of the PT2 theory as a correction (lrc), the resulting
lrc-XYG3 indeed improves over XYG3 notably \cite{zhang:2013A}. Moreover, the good performance of lrc-XYG3 holds for the whole binding curve.
A recent work fully relaxed 7 parameters in the xDH@B3LYP model (Eq.~\ref{eq:xdh@b3lyp}) and proposed the XYG7 method \cite{zhang:2021A}.
Fig.~\ref{fig:MB_dissociation} indicates that it is possible to accurately describe vdW interactions in the context of xDH@B3LYP without resorting
to the use of the explicit dispersion correction.

It was also found that the aforementioned underestimation of vdW interaction can be largely corrected if the Pople basis set of 6-311+G(3df,2p)
is employed \cite{zhang:2012B}. XYG3@6-311+G(3df,2p) and XYGJ-OS@6-311+G(3df,2p) perform well for the S22 set. The overall
MAEs are only about 10 meV for both methods \cite{zhang:2012B}. Recently, Chen and Xu extended the applicability of XYG3 to molecular crystals, which
was fulfilled with the aid of the eXtended ONIOM method (XO) with periodic boundary conditions (XO-PBC) \cite{xu:2020}. The X23 data set consists
of 23 molecular crystals dominated by hydrogen bondings, vdW interactions, and those of combined hydrogen bonding-vdW interactions \cite{tkatchenko:2013}.
Chen and Xu computed the lattice energies of these 23 molecular crystals, and reported a MAE of only 30 meV for the XYG3 method \cite{xu:2020}.

In addition to vdW \index{vdW} complexes, DHAs, RPA \index{RPA} and its variants have also been applied to chemically bonded molecules and 
crystalline solids. Interested readers may look into the literature for further details \cite{Harl/Schimka/Kresse:2010,zhang:2009A, Zhang/Xu:2014, 
	Paier/etal:2012, Ren/etal:2012b,Ren/etal:2013,zhang:2021C}.

\subsection{Surface adsorption}

An accurate description of atoms and molecules interacting with surfaces is the key to understand important physical and chemical processes 
such as heterocatalysis, molecular electronics, and corrosion. Molecules adsorbed on metal surfaces represent a particularly difficult situation, 
since quantitatively accurate results can only be obtained
if the approach is able to describe simultaneously the chemical bonding, vdW \index{vdW} interactions, metallic screening, and charge transfer processes.
The CO adsorption problem and the water hexamer adsorbed on the Cu(111) surface, to be discussed below, are two prominent examples which
require fifth-rung functionals to give a faithful description. 

\subsubsection{The ``CO adsorption puzzle''.}  
\begin{figure}[t]
 \centering
	\begin{picture}(400,155)(0,0)
	\put(10,150){(a)}
	\put(20,35){\includegraphics[width=0.3\textwidth]{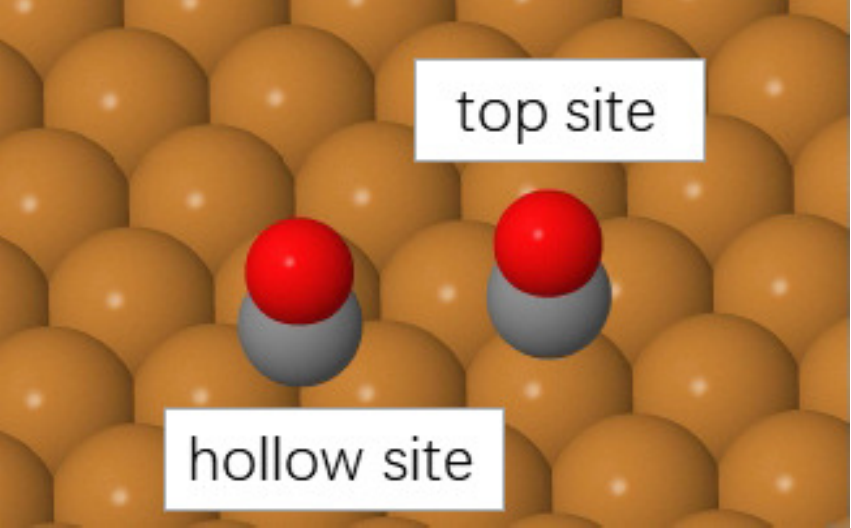}}
		\hskip 5mm
	\put(190,0){\includegraphics[width=0.45\textwidth]{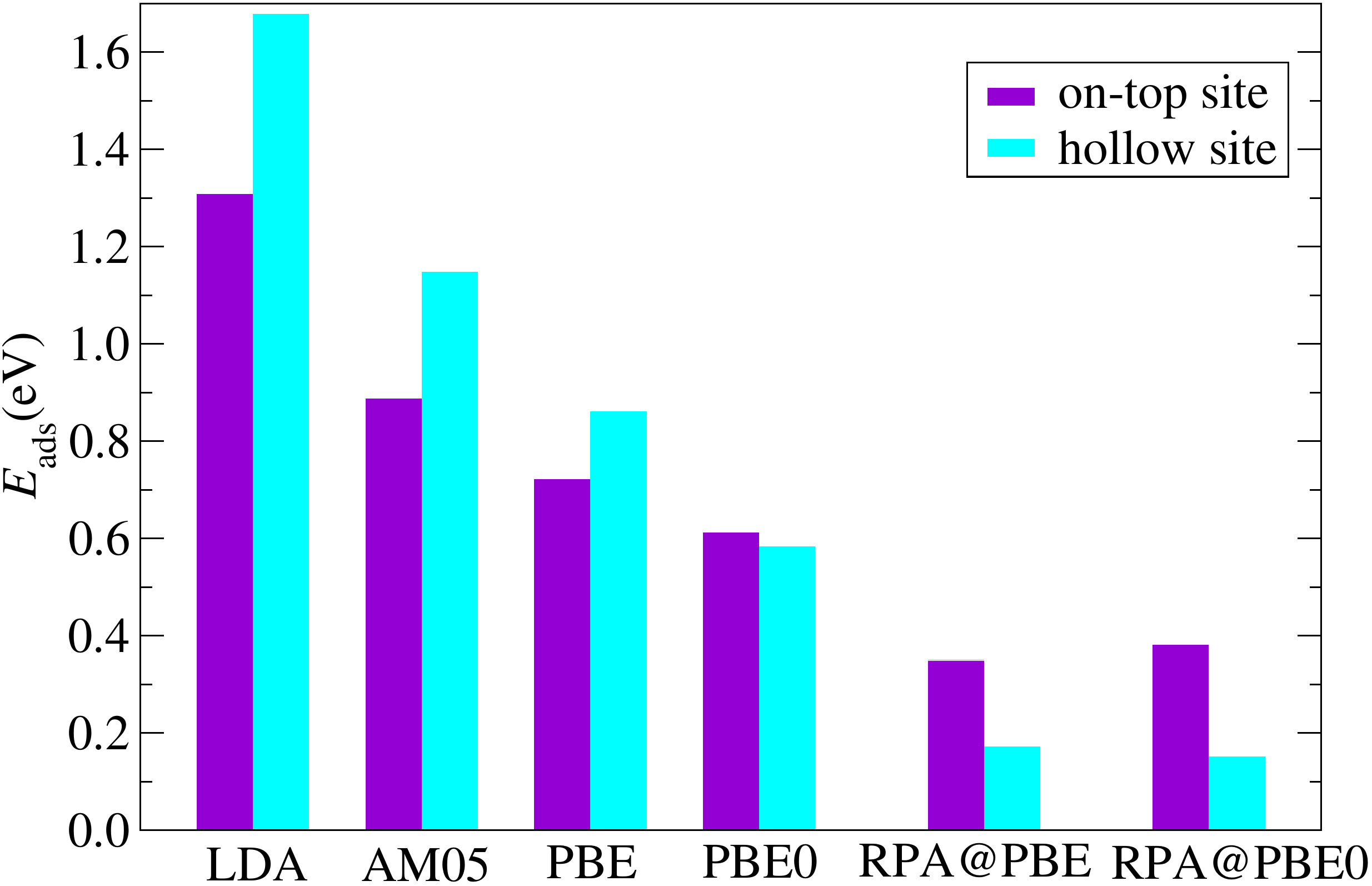}}
	\put(180,150){(b)}
	\end{picture}
	\caption{\label{Fig:CO_ads}(a): Schematic of CO adsorbed on the Cu(111) surface, with the on-top and hollow adsorption sites illustrated. 
	      (b): Adsorption energies for CO sitting on the on-top and fcc hollow sites 
               as obtained using LDA, AM05, PBE, PBE0, and RPA. RPA results are presented
	       for both PBE and PBE0 references. Calculations were done with the FHI-aims code. Adopted from Ref.~\cite{Ren/etal:2012b}.}
\end{figure}

This issue can be well illustrated by the ``CO adsorption puzzle", 
where LDA and several popular GGAs predict the wrong adsorption site for the 
CO molecule adsorbed on several noble/transition metal surfaces at low coverages \cite{Feibelman:2001}. For CO adsorbed on the Cu(111), 
Pt(111), and Rh(111) surfaces,
LDA and popular GGAs erroneously favor the threefold-coordinated hollow site (cf. Fig.~\ref{Fig:CO_ads}(a)), whereas experiments clearly show that the 
singly-coordinated on-top site is the energetically most stable site \cite{Steininger/Lehwald/Ibach:1982,Blackman/etal:1988}. 
This posed a severe challenge to the first-principles modeling of molecular adsorption problems and represents a key test example for the RPA-based methods.

In Ref.~\cite{Ren/etal:2009}, the performance of the RPA \index{RPA} method for CO adsorbed on Cu(111) was investigated. The computed RPA
adsorption energies for both the on-top and fcc (face centered cubic) hollow sites are presented in \ref{Fig:CO_ads}(b), together with the results from LDA, 
AM05 \cite{Armiento/Mattsson:2005}, PBE, and the hybrid PBE0 functional \cite{Perdew/Ernzerhof/Burke:1996}.
Figure~\ref{Fig:CO_ads} reveals what happens in the CO adsorption problem when climbing the Jacob's ladder \index{Jacob's ladder} in DFT \cite{Perdew/Schmidt:2001}, when going from the first two rungs (LDA and GGAs) to the fourth (hybrid functionals), and finally to the 5th-rung RPA \index{RPA} functional.
It can be seen that, along the way, the adsorption energies on both sites are reduced, while the magnitude of the reduction is bigger for the fcc hollow site. 
The correct energy ordering is already restored at the PBE0 level, but 
the adsorption energy difference between the two sites is vanishingly small. RPA \index{RPA} not only predicts the correct adsorption site, but also produces 
reasonable adsorption energy difference of 0.22 eV, consistent with experiments. The effect of the starting reference state on the calculated RPA results
has also been checked. In Ref.~\cite{Ren/etal:2009}, in addition to the commonly used RPA@PBE scheme, RPA calculations were also performed on top of the
hybrid functional PBE0. Figure~\ref{Fig:CO_ads} indicates that the small difference between RPA@PBE and RPA@PBE0 results is insignificant for understanding 
the ``CO adsorption puzzle".  Schimka \textit{et al.} extended the RPA benchmark studies of the CO adsorption problem to more surfaces 
\cite{Schimka/etal:2010}, and found that RPA is the only approach that gives both good adsorption energies and surface energies at the same time. 
GGAs and hybrid functionals at most yield either good surface energies, or adsorption energies, but not both. 
Following these works, RPA has subsequently been applied to the adsorption of small alkanes in Na-exchanged chabazite \cite{Goltl/Hafner:2011}, 
benzene on the Si(001) surface \cite{Kim/etal:2012}, graphene on the Ni(111) surface \cite{Mittendorfer/etal:2011,Olsen/etal:2011} 
and the Cu(111) and Co(0001) surfaces \cite{Olsen/etal:2011}.
In all these studies, RPA was demonstrated to be able to capture the delicate balance between chemical and dispersion interactions, and yields quantitatively
reliable results. We expect RPA to become an increasingly more important approach in surface science, 
with increasing computer power and more efficient implementations.

\subsubsection{The water hexamer on Cu(111) surfaces} 
\begin{figure}[t]
 \centering
	\includegraphics[width=0.8\textwidth]{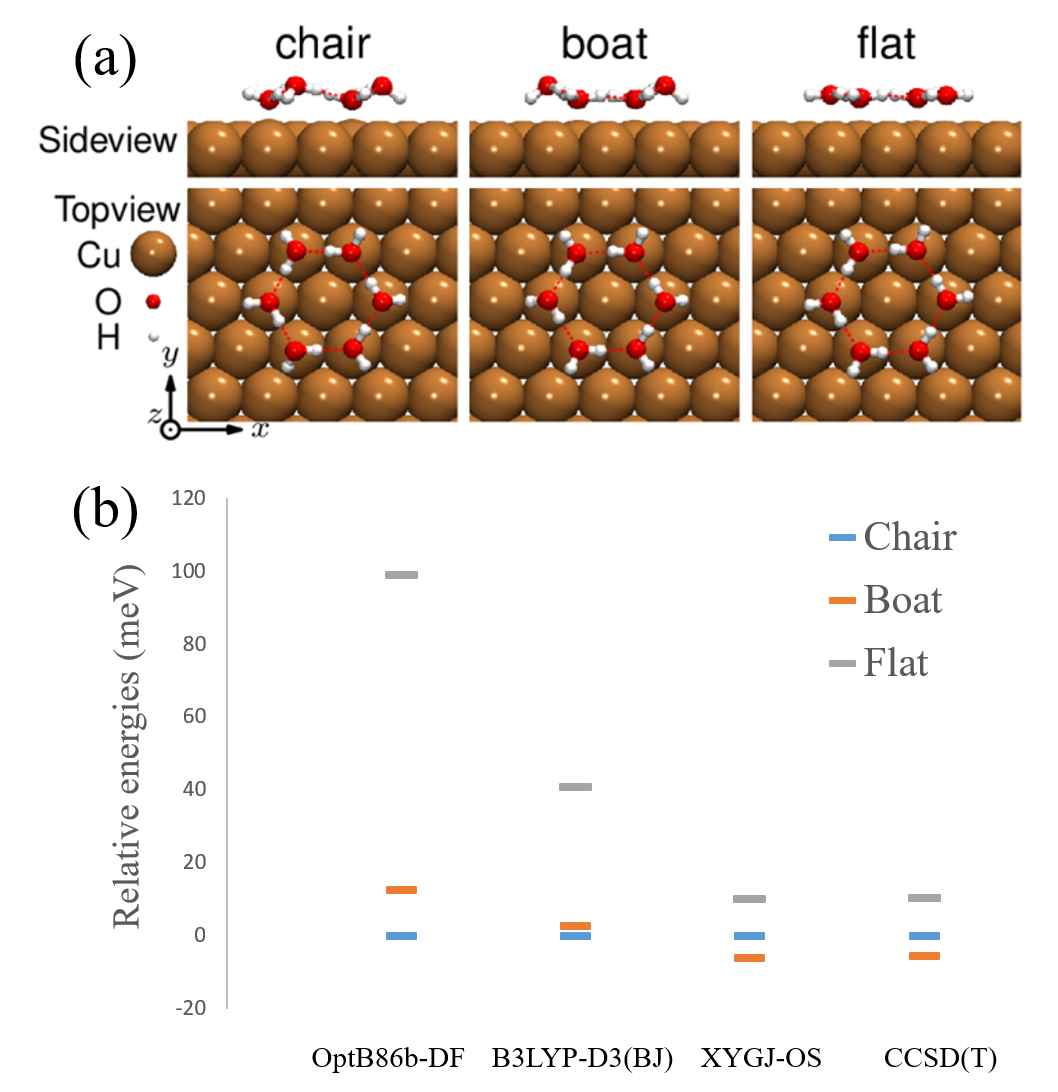}
	\caption{\label{Fig:water_ads}(a): Schematic of (H$_2$O)$_6$ adsorbed on the Cu(111) surface with ``chair'', ``boat'', and ``flat'' configurations. 
	(b): The relative adsorption energies of three kinds of configuration calculated by OptB86b-DF, B3LYP-D3(BJ), XYGJ-OS, and CCSD(T). 
	The adsorption energies of the ``chair'' configuration are set to zero.
    Adopted from Ref.~\cite{duan:2020}.}
\end{figure}

is a more complicated adsorption system, which also imposes a great challenge in computational 
materials science \cite{duan:2020}. As shown in Fig.~\ref{Fig:water_ads}, the conventional DFAs, including OptB86b-DF and B3LYP-D3(BJ), predict the
chair configuration to be the most stable adsorption structure of the water hexamer on Cu(111) surfaces. However, the scanning tunneling
microscope (STM) image from experiment supported a more symmetric configuration, and led to the assignment of the flat
configuration to be the in situ configuration \cite{michaelides:2007}. Recently, Duan et. al. revisited this problem by using the XYG3-type DHA XYGJ-OS
and advanced wave-function method CCSD(T) \cite{duan:2020}. With a more accurate description of hydrogen bond interaction, XYGJ-OS and CCSD(T)
changes the configuration energy ordering. Together with the STM simulation, these authors identified the boat configuration to be the most stable one, solving 
the aforementioned discrepancy between experiment and simulation.

\subsection{Structural phase transition}
\subsubsection{The $\alpha$-$\gamma$ phase transition of Ce.}
\begin{figure}[t]
      \vskip 5mm
      \centering
      \includegraphics[width=0.45\textwidth]{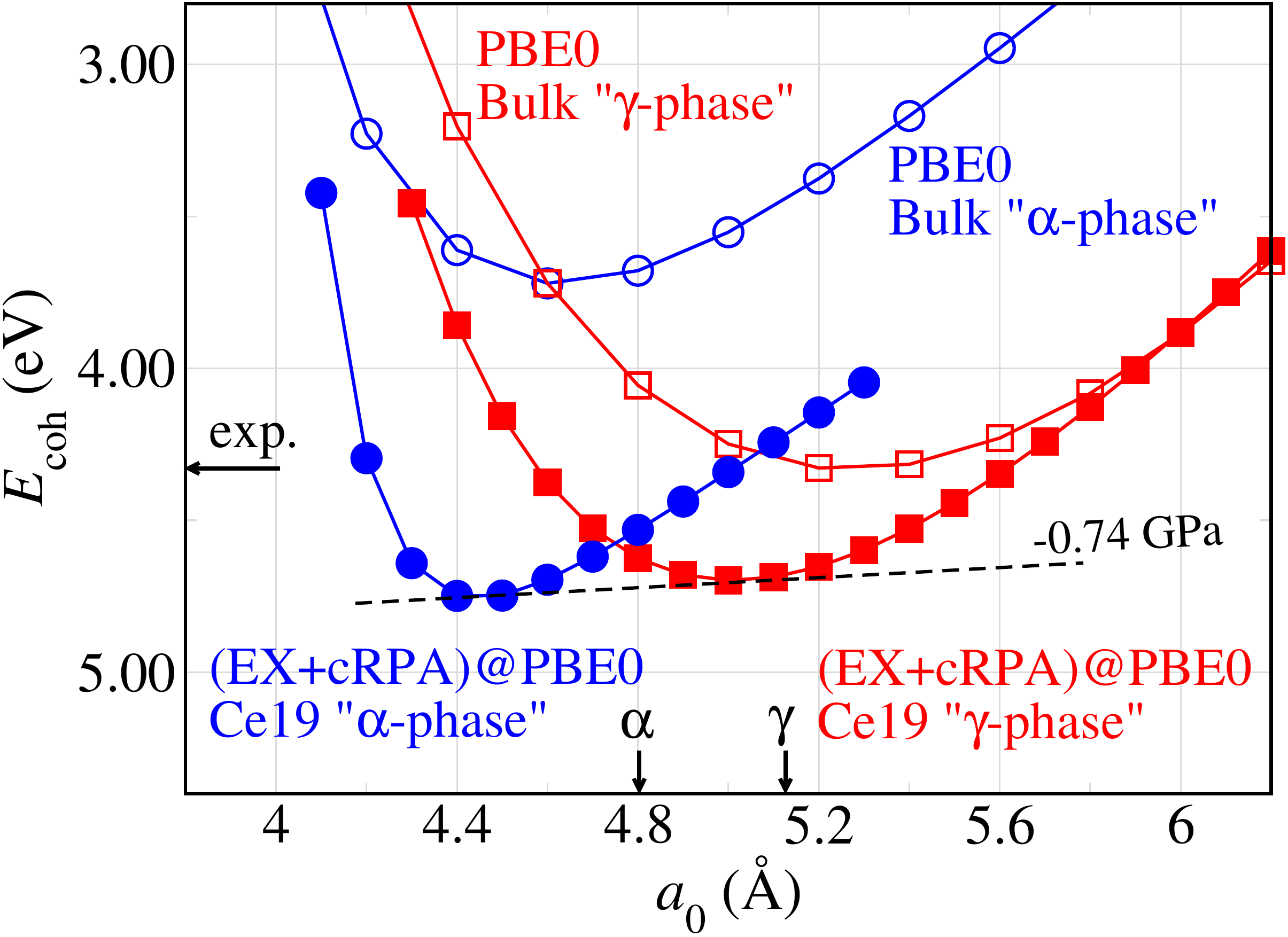}
	\caption{\label{fig:Ce_RPA} Calculated PBE0 and RPA@PBE0 [RPA denoted here as exact-exchange plus RPA correlation (EX+cRPA)] 
	 cohesive energy ($E_{\mathrm{coh}}$) as a function of the lattice constant $a_0$ for the two electronic states based on self-consistent PBE0 solutions. 
	 The correction of RPA with respect to the PBE0 cohesive energies was done for a 19-atom
	 fcc-cerium cluster. The dashed line illustrates the Gibbs construction for the transition pressure from the $\alpha$ phase to the $\gamma$ phase. 
	 Error on the energy axes indicates the experimental cohesive energy of the $\alpha$ phase. Adopted from Ref.~\cite{Casadei/etal:2012}.}
\end{figure}

The $f$-electron materials, which contain rare-earth or actinide elements, pose a great challenge to first-principles approaches. A prototypical example of 
$f$-electron system is the Ce metal, which has an intriguing iso-structural $\alpha$-$\gamma$ phase transition, accompanied by a drastic volume collapse 
(as large as $15\%$ at ambient pressure and zero temperature). The two phases are characterized by distinct spectroscopic and magnetic properties.
Various theoretical approaches, including LDA plus dynamical mean-field theory (LDA+DMFT) have been employed to study this system 
\cite{Held/etal:2001,Amadon/etal:2006}.
DFT within its local and semilocal approximations is unable to describe the two phases of Ce. In fact, due to the strong delocalization error in these functionals,
the localized nature of the $f$-electrons in the $\gamma$ phase cannot be properly described; only the $\alpha$ phase is described with some confidence
within LDA/GGAs, although not at a quantitative level. 

It was shown by Casadei \textit{et al.} \cite{Casadei/etal:2012,Casadei/etal:2016} that, remarkably, hybrid functionals 
like PBE0 and HSE can yield two self-consistent solutions,
with distinct lattice constants, cohesive energies, electronic band structures, local magnetic moments, and different degrees of $f$-electron localization.
These two solutions can be reasonably associated with the phases of the Ce metal. However, the energetic ordering of the $\alpha$-like and $\gamma$-like
phases produced by PBE0 is not consistent with the experimental situation where the $\alpha$ phase is energetically more stable at low temperature and
ambient pressure. Adding RPA corrections on top of the PBE0 cohesive energies for the two solutions, the energy ordering is reversed, and the 
$\alpha$-like solution becomes energetically more stable. The transition pressure of the two phases given by the Gibbs construction is consistent with
the experimental value.

\subsubsection{The fcc-hcp phase transition of Ar.}

The face-centered cubic (fcc) and hexagonal closed packed (hcp) structures of the rare-gas crystals have very close energies, but at ambient pressures 
the fcc crystal structure is preferred and there is a phase transition from fcc to hcp at very high pressures. Because of the tiny energy difference between the two phases, it is exceedingly difficult to capture this difference computationally and consequently explain these behaviors. Conventional approximations of DFT do not have the adequate accuracy to describe the system. In the literature, the problem was usually treated by the cluster expansion approach, whereby the two-body, three-body, and four-body potentials are obtained either from empirical models or quantum chemistry  coupled cluster 
calculations \cite{PhysRevLett.79.1301,PhysRevB.62.5482,PhysRevB.60.7905,Schwerdtfeger2016Towards}. However, this type of approach is difficult to use for non-experts and not suitable for routine calculations.

Thus, it is highly interesting to check how fifth-rung functionals perform for discerning the fcc and hcp structures of the rare-gas crystals. 
In Ref.~\cite{Yang/Ren:2022}, focusing on the Ar crystal, the authors found that the correct energy ordering between the fcc and hcp phases at ambient pressure can be obtained if 1) the RPA or RPA+rSE method is employed and 2) the fcc and hcp crystal structures are computationally treated on an equal footing (i.e., invoking the same computational supercell and $\bfk$ grid mesh). In contrast with what was previously proposed, it was found that the electronic correlation effect plays a decisive role in determining the relative stability of the two phases, whereas the zero point energy effect is secondary.

\begin{figure}[t]
\centering
\includegraphics[width=0.65\textwidth,clip]{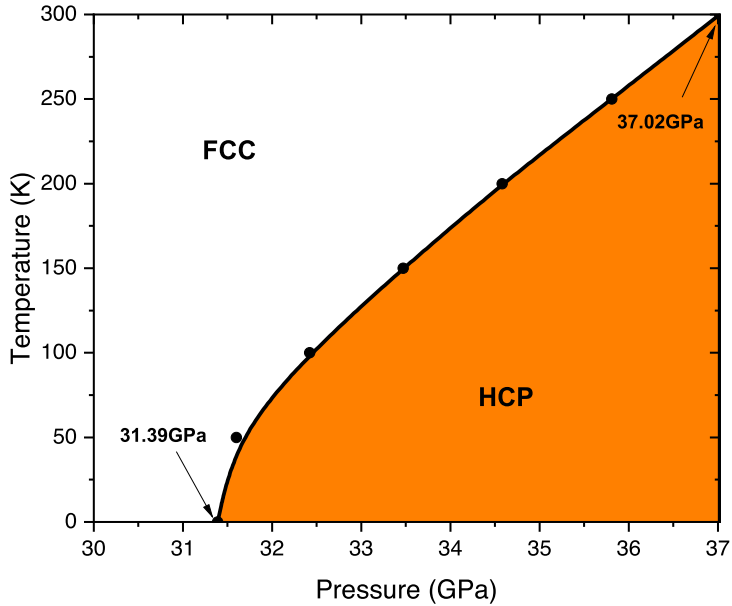}
\caption{\label{Fig:T-P_diagram} The $T-P$ diagram of the Ar crystal, where the phase boundary between the fcc and hcp phases are determined by the
Gibbs free energy. Calculations are done using the 6-atom supercell model at several successive temperatures, i.e.,  $0$, 50, 100, $\cdots$, $300$ K . 
Both electronic (RPA+rSE) and phonon (PBE) parts of the Gibbs free energy are extrapolated to the complete basis set limit. Adopted 
from Ref.~\cite{Yang/Ren:2022}.}
\end{figure}

Furthermore, by complementing the RPA+rSE electronic ground-state energy with lattice vibrational energies at finite temperatures, one can compute the Helmholtz free energy and derive the pressure-volume ($P$-$V$) curve at a given temperature. The calculated $P$-$V$ curve is in excellent agreement with the experimental measurements \cite{Yang/Ren:2022}, especially in the high-pressure regime, which has been considered a big challenge for theoretical approaches. Finally, by computing the Gibbs free energies for both phases, it is possible to determine a temperature-pressure ($T$-$P$) phase diagram for the Ar crystal, which signifies the rough temperature and pressure range where the fcc phase can be transformed to the hcp phase. Such a $T$-$P$ phase
diagram is shown in Fig.~\ref{Fig:T-P_diagram}.

\subsubsection{Rutile v.s.~Anatase phases for TiO$_2$.}  
\begin{figure}[t]
\centering
\includegraphics[width=0.65\textwidth,clip]{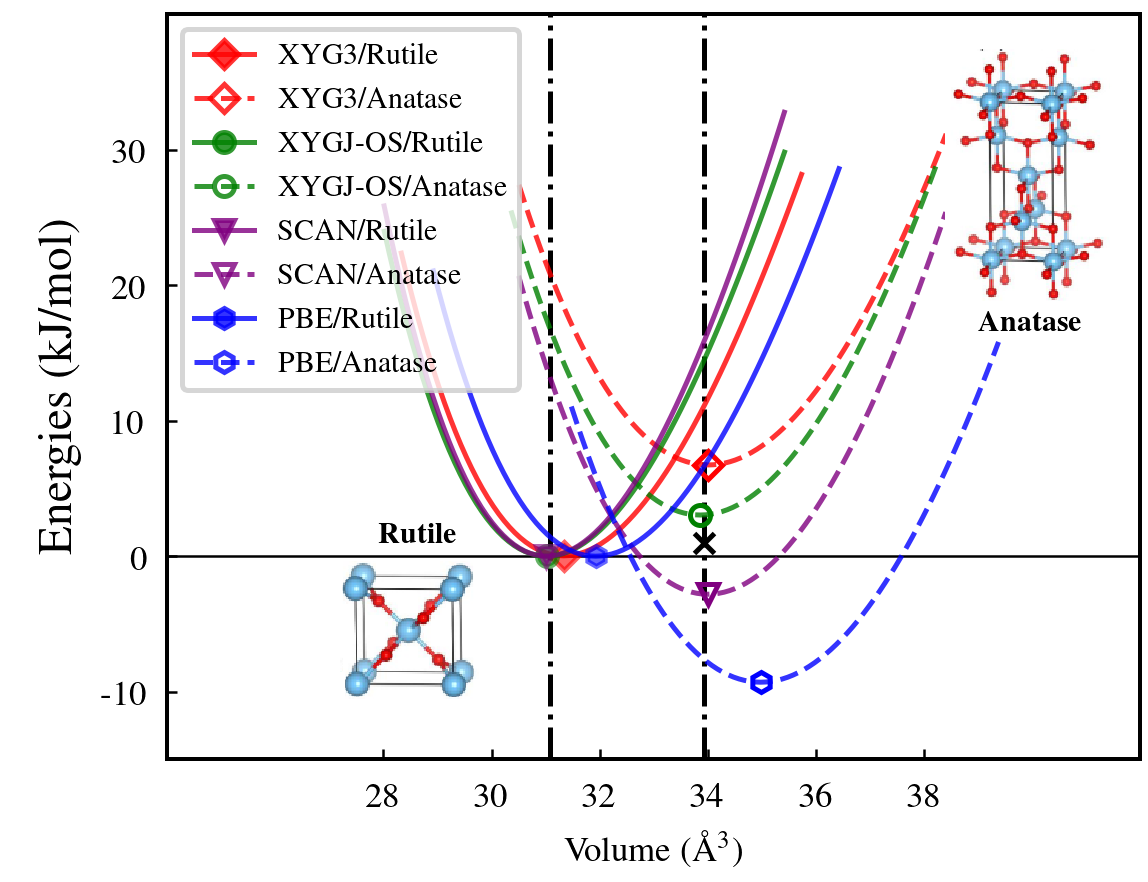}
\caption{\label{Fig:tio2} Energies (kJ/mol) of TiO$_2$ with rutile and anatase structures as a function of the volume of the primitive cell 
	(\AA$^3$). The energy zero for all DFAs is the equilibrium energy of rutile TiO$_2$ calculated by the method itself. Optimized volume 
	and the corresponding cohesive energies are marked for various DFAs, while the experimental values (with zero-point energy contribution excluded) 
        are given by a black ``X''. Adopted from Ref.~\cite{zhang:2021C}.}
\end{figure}

Recently, PT2-based DHAs have been applied to study the relative stability of TiO$_2$,
which is also a challenging case in materials science. It was clear from the experiment that the rutile phase is the most stable phase at
T=0 K \cite{sun:2019}. However, as shown in Fig.~\ref{Fig:tio2}, most of the popular DFAs, including PBE and SCAN, predict the anatase phase to be
more stable than rutile. Zhang et.~al.~attributed this issue to the self-interaction error in conventional DFAs \cite{sun:2019}. 
They thus suggested to use the empirical DFT+U approach to effectively reduce SIE for PBE and SCAN. 
However, in order to obtain the correct phase ordering, a very large U is needed (6 eV for PBE and 2 eV for SCAN), which notably deteriorates
the description of the structure parameters. On the other hand, Cui et.~al. suggested that such kind of error should be attributed to 
the error in the (semi)local correlation approximations. They found that the correct energy ordering can be obtained by using the RPA
method \cite{jiang:2016}. Fig.~\ref{Fig:tio2} shows that the two XYG3-type DHAs, including XYG3 and XYGJ-OS, correctly predict the rutile TiO$_2$ 
to be the ground phase with a satisfactory description of the optimized equilibrium volume.

In addition to the three systems discussed above, 
the capability of RPA to capture the delicate energy difference between competing phases or polymorphs has also been demonstrated for 
Iron disulfide (FeS$_2$), a potentially interesting system for photovoltaic and photoelectrochemical applications. This material turns out to be 
rather challenging, since popular conventional DFAs fail to produce the relative stability of its two
polymorphs: pyrite and marcasite, with the latter artificially stabilized.  It was demonstrated by Zhang \textit{et al.} \cite{Zhang/Cui/Jiang:2018}
that RPA, regardless of its reference state, can correctly restore the energy ordering of the two polymorphs of FeS$_2$. These authors further unravel
that the fact that RPA tends to stabilize the pyrite polymorph is due to its smaller KS band gap, resulting in a large RPA correlation energy
as compared to the the marcasite polymorph. This observation is consistent with the case of the Ce metal \cite{Casadei/etal:2012}, where the more 
metallic $\alpha$-phase is stabilized within the RPA. Another successful application of this kind is that the tiny energy difference between the
two allotropes of carbon, graphite and diamond, can be reliably described by RPA, with the correct prediction that graphite is energetically slightly
lower than diamond \cite{Lebegue/etal:2010}. Similarly, the thermodynamic stability of the most common polymorphs of the bulk BN has been resolved
by Cazorla and Gould \cite{Cazorla/Gould:2019} at the level of RPA. A more systematic benchmark study of the performance of RPA and several beyond-RPA methods
for predicting the transition pressure of structural phase transitions of a set of bulk materials have recently been reported in Ref.~\cite{Sengupta/etal:2018}.

Other types of materials science problems to which fifth-rung functionals have been applied include
layered compounds \cite{Marini/Gonzalez/Rubio:2006,Bjoerkman/etal:2012} and defect systems \cite{Bruneval:2012,Kaltak/Klimes/Kresse:2014}. 
We shall not elaborate on these types of applications here due to limited space, and interested readers may look into the original literature for details.
In summary, there are ample evidence that fifth-rung DFAs perform well in capturing delicate energy differences in materials, and fix some
of the qualitative failures of more conventional approaches. So far, RPA-based method and DHAs are mostly applied by
different groups to different materials science problems.  A systematic comparison of the performance of these two types of fifth-rung functionals 
for a well-defined set of benchmark systems still needs to be done in the near future.

\section{Recent developments}

The field of fifth-rung functional development and applications represents a rapidly evolving branch of computational electronic structure methods. 
Noteable progress has been achieved in several directions within the last few years, which we would like to briefly recapitulate here.
  \begin{enumerate}
    \item Force calculations of RPA and DHAs. Analytical gradients of the RPA \index{RPA} total energy with respect to the atomic positions have been 
		computed within both the atomic orbital basis \cite{Rekkedal/etal:2013,Burow/etal:2014,Mussard/etal:2014,Beuerle/Ochsenfeld:2018,
		Chedid/Jocelyn/Eshuis:2021,Tahir/etal:2022} and plane-wave basis \cite{Ramberger/etal:2017} frameworks. 
		Corresponding implementation for DHAs can be found mainly with the atomic orbital basis framework \cite{su:2013}.
		This allows to compute atomic forces and relax structures at the levels of RPA and DHAs, which is a long-sought goal of the 
		electronic-structure community. 
	    Moreover, it is now possible to calculate vibrational frequencies \cite{Burow/etal:2014} and phonon spectra based on
        the RPA force constant \cite{Ramberger/etal:2017}, and even molecular dynamics simulations can be carried out based on the RPA
	    force \cite{Ramberger/etal:2017}.
        These advancements greatly enhanced the capability of these advanced methods in computational chemistry and materials science.
   \item Low-scaling implementations for RPA and DHAs.
      Another noteworthy development is several low-scaling -- ranging from linear scaling to cubic scaling -- algorithms for RPA 
	  \cite{Neuhauser/etal:2013,Moussa:2014,Kaltak/Klimes/Kresse:2014,Kallay:2015,Wilhelm/etal:2016,Graf/etal:2018,Luenser/Schurkus/Ochsenfeld:2017,Lu/Thicke:2017,Duchemin/Blase:2019} and DHAs\cite{zhang:2021D,martin:2022} have been designed and implemented. 
      This paves the way for applying fifth-rung functionals to large-sized and complex materials that are unaccessible in the past.
   \item Particle-particle RPA. 
      The above-discussed RPA is represented by ring diagrams and called particle-hole RPA (phRPA) in the literature. In addition to phRPA, another type of RPA, 
      consisting of an infinite summation of ladder diagrams, has also been discussed in the nuclear physics literature \cite{Ring/Schuck:1980}. 
      This type of RPA is referred to as particle-particle RPA and has recently been brought into the attention of 
      electronic structure community \cite{Aggelen/etal:2013,Peng/Steinmann/etal:2013,Yang/Aggelen/Steinmann/etal:2013,Scuseria/Henderson/Bulik:2013}. 
      Benchmark calculations show that ppPRA carries interesting features that are not present in phRPA. Attempts for combining the two types of RPA in 
      one framework has been made both in a range-separated manner \cite{Shepherd/Henderson/Scuseria/:2014} and globally \cite{Tahir/Ren:2019}.
      However, it seems that merging the two RPA channels into one united theory is a highly nontrivial task \cite{Tahir/Ren:2019}.
   \item Self-consistent RPA and DHAs in generalized Kohn-Sham framework. As mentioned before, the majority of practical RPA and DHAs calculations 
	   are done in a 
	   post-processing fashion, using orbitals and orbital energies generated from a preceding DFA calculation.
	   The importance of the rSE \index{rSE} contribution indicates
       that commonly used semi-local DFAs are not optimal starting points for RPA calculations. Thus running the calculations of RPA and DHAs 
	   in a self-consistent way is highly desirable. However, for orbital-dependent functionals like RPA and DHAs, the criterion for
	   ``self-consistency" is not uniquely defined. 
       In the RPA case, the optimized effective potential (OEP) scheme \cite{Verma/Bartlett:2012,Bleiziffer/Hesselmann/Goerling:2013,Hellgren/etal:2015} 
	   is a well-defined self-consistency procedure, but the obtained results for binding energies are not necessarily better than than the
         perturbative scheme. 
	   Similar observation was found for DHAs\cite{head-gordon:2013,head-gordon:2018}.
	   Most recently, self-consistent RPA schemes are developed by
       Jin \textit{et al.} within generalized OEP framework \cite{Jin/etal:2017} and by Voora \textit{et al.} within generalized KS framework \cite{Voora/etal:2019,Yu/Furche_etal:2021}. 
		  The two schemes differ in details, but the rSE \index{rSE} contribution is captured in both schemes. Initial results obtained from these schemes look
       very promising and there is much to explore along this direction.
    \item More beyond-RPA schemes to improve the accuracy. In addition to the beyond-RPA schemes already discussed in Sec.~\ref{sec:beyond_rpa},
       one most recent development by Zhang and Xu \cite{Zhang/Xu:2019} is to introduce a spin-pair distinctive algorithm in the ACFDT context, whereby
       the contributions of same-spin and opposite-spin pairs to the correlation energy are separated. By scaling the contributions for
	   two types of spin pairs differently, one can achieve a simultaneous attenuation of both self-interaction errors and static correlation errors. 
	   Similar to the power series approximation of G\"{o}rling and coauthors \cite{Erhard/etal:2016,Goerling:2019}, the spin-pair distinctive
	   formalism of Zhang and Xu \cite{Zhang/Xu:2019} is particularly successful in dealing with systems with multi-reference characters.
  \end{enumerate}

\section{Summary}

In this review, we discussed the basic concept, the theoretical formulation, the implementation algorithm, the prototypical applications, as well as
the recent developments of PT2-based DHAs and RPA-based methods. We expect that these fifth-rung functionals will have an ever-increasing impact 
on computational materials science, and become the  mainstream methods in the near future. Since this is an actively developing field, 
and the number of papers is quickly growing, 
it is well possible some important developments are not covered in our discussion. The purpose of this manuscript is to inform the readers about
the overall status of this field, and stimulate more works on the development and applications of fifth-rung functional methodology.

\section{Acknowledgements}
We are grateful to Matthias Scheffler, Xin Xu, Patrick Rinke and other colleagues for great collaborations and illustrating discussions 
on the works presented in this review.
The authors acknowledge the financial support of National Natural Science Foundation of China (Grant Nos. 12134012, 12188101, 21973015, and 22125301).
This work was also supported by the funding of Innovative research team of high-level local universities in Shanghai and a key laboratory program
of the Education Commission of Shanghai Municipality (ZDSYS14005).
X.R. acknowledges the financial support of  the Max Planck Partner Group for
\textit{Advanced Electronic Structure Methods}.

\clearpage
\bibliography{./CommonBib.bib}


\end{document}